\documentclass[12pt,a4paper]{article}
\usepackage{ifthen} 
\newboolean{articletitles}
\setboolean{articletitles}{true} 
\newboolean{uprightparticles}
\setboolean{uprightparticles}{false} 
\newboolean{inbibliography}
\setboolean{inbibliography}{false} 
\pdfoutput=1

\usepackage{microtype}
\usepackage{xspace} 

\usepackage{comment}

\usepackage{graphicx}  
\usepackage{color}
\usepackage{colortbl}
\graphicspath{{./figs/}} 

\usepackage{amsmath} 
\usepackage{amssymb}
\usepackage{amsfonts}
\usepackage{mathrsfs,bm,url,color}
\usepackage{upgreek} 
\usepackage{blindtext} 

\usepackage{hyperref}    
\usepackage[all]{hypcap} 

\def\ux85 {\mbox{UX85}\xspace}

\ifthenelse{\boolean{uprightparticles}}%
{

 \def\PDelta      {\ensuremath{\Delta}\xspace}                 
 \def\PXi      {\ensuremath{\Xi}\xspace}                 
 \def\PLambda      {\ensuremath{\Lambda}\xspace}                 
 \def\PSigma      {\ensuremath{\Sigma}\xspace}                 
 \def\POmega      {\ensuremath{\Omega}\xspace}                 
 \def\PUpsilon      {\ensuremath{\Upsilon}\xspace}                 
 
 \def\PB      {\ensuremath{\mathrm{B}}\xspace}                 
                  
 \def\PD      {\ensuremath{\mathrm{D}}\xspace}

 \def\PK      {\ensuremath{\mathrm{K}}\xspace}

 \def\Pc      {\ensuremath{\mathrm{c}}\xspace}

 \def\Pi      {\ensuremath{\mathrm{i}}\xspace}

}
{

 \mathchardef\PDelta="7101
 \mathchardef\PXi="7104
 \mathchardef\PLambda="7103
 \mathchardef\PSigma="7106
 \mathchardef\POmega="710A
 \mathchardef\PUpsilon="7107
                  
 \def\PB      {\ensuremath{B}\xspace}                 
                  
 \def\PD      {\ensuremath{D}\xspace}

 \def\PK      {\ensuremath{K}\xspace}

 \def\Pc      {\ensuremath{c}\xspace}

 \def\Pi      {\ensuremath{i}\xspace}

}

\def\ellell     {\ensuremath{\ell^+ \ell^-}\xspace}

\def\cquark    {\ensuremath{\Pc}\xspace}

\def\piz   {\ensuremath{\pion^0}\xspace}

\def\pip   {\ensuremath{\pion^+}\xspace}
\def\pim   {\ensuremath{\pion^-}\xspace}

\def\kaon  {\ensuremath{\PK}\xspace}
  \def\Kbar  {\kern 0.2em\overline{\kern -0.2em \PK}{}\xspace}

\def\Kz    {\ensuremath{\kaon^0}\xspace}
\def\Kzb   {\ensuremath{\Kbar^0}\xspace}
\def\KzKzb {\ensuremath{\Kz \kern -0.16em \Kzb}\xspace}
\def\Kp    {\ensuremath{\kaon^+}\xspace}
\def\Km    {\ensuremath{\kaon^-}\xspace}

\def\KpKm  {\ensuremath{\Kp \kern -0.16em \Km}\xspace}
\def\KS    {\ensuremath{\kaon^0_{\rm\scriptscriptstyle S}}\xspace}

\def\Kstarp  {\ensuremath{\kaon^{*+}}\xspace}

\def\Dbar    {\kern 0.2em\overline{\kern -0.2em \PD}{}\xspace}
\def\D       {\ensuremath{\PD}\xspace}

\def\Dz      {\ensuremath{\D^0}\xspace}
\def\Dzb     {\ensuremath{\Dbar^0}\xspace}
\def\DzDzb   {\ensuremath{\Dz {\kern -0.16em \Dzb}}\xspace}
\def\Dp      {\ensuremath{\D^+}\xspace}
\def\Dm      {\ensuremath{\D^-}\xspace}

\def\DpDm    {\ensuremath{\Dp {\kern -0.16em \Dm}}\xspace}

\def\B       {\ensuremath{\PB}\xspace}
\def\Bbar    {\ensuremath{\kern 0.18em\overline{\kern -0.18em \PB}{}}\xspace}

  \def\Y#1S{\ensuremath{\PUpsilon{(#1S)}}\xspace}

\def\L {\ensuremath{\PLambda}\xspace}
\def\Lbar {\ensuremath{\kern 0.1em\overline{\kern -0.1em\PLambda}}\xspace}

\def\Lc      {\ensuremath{\L^+_\cquark}\xspace}

\def\BF         {{\ensuremath{\cal B}\xspace}}

\def\BR         {\BF}

\def\to                 {\ensuremath{\rightarrow}\xspace}

\def\CP                {\ensuremath{C\!P}\xspace}

\def\AT#1     {\ensuremath{A_{\mathrm{T}}^{#1}}\xspace}           

\def\C#1      {\ensuremath{\mathcal{C}_{#1}}\xspace}                       
\def\Cp#1     {\ensuremath{\mathcal{C}_{#1}^{'}}\xspace}                    
\def\Ceff#1   {\ensuremath{\mathcal{C}_{#1}^{\mathrm{(eff)}}}\xspace}        
\def\Cpeff#1  {\ensuremath{\mathcal{C}_{#1}^{'\mathrm{(eff)}}}\xspace}       
\def\Ope#1    {\ensuremath{\mathcal{O}_{#1}}\xspace}                       
\def\Opep#1   {\ensuremath{\mathcal{O}_{#1}^{'}}\xspace}                    

\newcommand{\tev}{\ensuremath{\mathrm{\,Te\kern -0.1em V}}\xspace}
\newcommand{\gev}{\ensuremath{\mathrm{\,Ge\kern -0.1em V}}\xspace}
\newcommand{\mev}{\ensuremath{\mathrm{\,Me\kern -0.1em V}}\xspace}
\newcommand{\kev}{\ensuremath{\mathrm{\,ke\kern -0.1em V}}\xspace}
\newcommand{\ev}{\ensuremath{\mathrm{\,e\kern -0.1em V}}\xspace}
\newcommand{\gevc}{\ensuremath{{\mathrm{\,Ge\kern -0.1em V\!/}c}}\xspace}
\newcommand{\mevc}{\ensuremath{{\mathrm{\,Me\kern -0.1em V\!/}c}}\xspace}
\newcommand{\gevcc}{\ensuremath{{\mathrm{\,Ge\kern -0.1em V\!/}c^2}}\xspace}
\newcommand{\gevgevcccc}{\ensuremath{{\mathrm{\,Ge\kern -0.1em V^2\!/}c^4}}\xspace}
\newcommand{\mevcc}{\ensuremath{{\mathrm{\,Me\kern -0.1em V\!/}c^2}}\xspace}

\def\gsim{{~\raise.15em\hbox{$>$}\kern-.85em
          \lower.35em\hbox{$\sim$}~}\xspace}
\def\lsim{{~\raise.15em\hbox{$<$}\kern-.85em
          \lower.35em\hbox{$\sim$}~}\xspace}

\def\pt         {\mbox{$p_{\rm T}$}\xspace}

\def\tell1  {TELL1\xspace}
\def\ukl1   {UKL1\xspace}

\newcommand{\eg}{\mbox{\itshape e.g.}\xspace}

\def\invab   {\ensuremath{\mbox{\,ab}^{-1}}\xspace}

\def\nut        {\ensuremath{\nu_\tau}\xspace}
\def\nub        {\ensuremath{\overline{\nu}}\xspace}

\def\nueb       {\ensuremath{\nub_e}\xspace}

\def\numb       {\ensuremath{\nub_\mu}\xspace}

\newcommand{\tauknu}   {\ensuremath{ \tau^{-} \to K^{-} \nut}\xspace}
\newcommand{\taupinu}   {\ensuremath{ \tau^{-} \to \pi^{-} \nut}\xspace}
\newcommand{\tauenu}   {\ensuremath{ \tau^{-} \to e^{-} \nueb \nut}\xspace}
\newcommand{\taumunu}  {\ensuremath{ \tau^{-} \to \mu^{-} \numb \nut}\xspace}
\newcommand{\BFKmutwo}      {\ensuremath{\BR(K^- \to \mu^- \numb)}\xspace}
\newcommand{\BFpimutwo}    {\ensuremath{\BR(\pi^- \to \mu^- \numb)}\xspace}
\newcommand{\BFtautoknu}    {\ensuremath{\BR(\tauknu)}\xspace}
\newcommand{\BFtautopinu}    {\ensuremath{\BR(\taupinu)}\xspace}

\newcommand{\BRtautomunu}    {\ensuremath{\frac{\BR(\taumunu)}{\BR(\tauenu)} }\xspace}

\usepackage{cite} 
\usepackage{mciteplus}
\usepackage{overpic}
\usepackage{relsize}

\usepackage{subfig} 

\usepackage[noabbrev,capitalise]{cleveref}

\usepackage{floatrow}

\newfloatcommand{capbtabbox}{table}[][\FBwidth]

\usepackage{caption}
\usepackage{longtable}
\usepackage{ifpdf}

\newcommand{\re}[2][()] {\ifthenelse{\equal{#1}{()}}{{\ensuremath{{\rm \, Re}}\left(#2\right)}}
                                                    {{\ensuremath{{\rm \, Re}}\left[#2\right]}}}
\newcommand{\im}[2][()] {\ifthenelse{\equal{#1}{()}}{{\ensuremath{{\rm \, Im}}\left(#2\right)}}
                                                    {{\ensuremath{{\rm \, Im}}\left[#2\right]}}}

\definecolor{orange}{rgb}{1,0.5,0}

\usepackage{booktabs}
\usepackage{rotating}
\usepackage{multirow}

\def\piz   {\ensuremath{\pi^0}\xspace}

\def\pip   {\ensuremath{\pi^+}\xspace}

\def\pim   {\ensuremath{\pi^-}\xspace}

\def\kaon  {\ensuremath{K}\xspace}
\def\Kbar  {\kern 0.2em\overline{\kern -0.2em K}{}\xspace}

\def\Kz    {\ensuremath{K^0}\xspace}
\def\Kzb   {\ensuremath{\Kbar^0}\xspace}
\def\KzKzb {\ensuremath{\Kz \kern -0.16em \Kzb}\xspace}
\def\Kp    {\ensuremath{K^+}\xspace}
\def\Km    {\ensuremath{K^-}\xspace}

\def\KpKm  {\ensuremath{\Kp \kern -0.16em \Km}\xspace}
\def\KS    {\ensuremath{K^0_{\scriptscriptstyle\rm S}}\xspace}

\def\D       {\ensuremath{D}\xspace}
\def\Dbar    {\kern 0.2em\overline{\kern -0.2em D}{}\xspace}

\def\Dz      {\ensuremath{D^0}\xspace}
\def\Dzb     {\ensuremath{\Dbar^0}\xspace}
\def\DzDzb   {\ensuremath{\Dz {\kern -0.16em \Dzb}}\xspace}
\def\Dp      {\ensuremath{D^+}\xspace}
\def\Dm      {\ensuremath{D^-}\xspace}

\def\DpDm    {\ensuremath{\Dp {\kern -0.16em \Dm}}\xspace}

\def\invab   {\ensuremath{\mbox{\,ab}^{-1}}\xspace}

\newcommand\snowmass{\begin{center}\rule[-0.2in]{\textwidth}{0.01in}\\\rule{\textwidth}{0.01in}\\
\vskip 0.1in Submitted to the  Proceedings of the US Community Study\\ 
on the Future of Particle Physics (Snowmass 2021)\\ 
\rule{\textwidth}{0.01in}\\\rule[+0.2in]{\textwidth}{0.01in} \end{center}}

\def\babar{\mbox{\slshape B\kern-0.1em{\smaller A}\kern-0.1em
    B\kern-0.1em{\smaller A\kern-0.2em R}}}

\floatstyle{plaintop}
\restylefloat{table}

\begin{document}
\renewcommand{\thefootnote}{\fnsymbol{footnote}}
\setcounter{footnote}{1}
\begin{titlepage}

\snowmass 
\vspace*{1.5cm}

{\bf\boldmath\huge
\begin{center}
Snowmass White Paper: \\Belle II physics reach and plans for the next decade and beyond
\end{center}
}

\vspace*{0.5cm}

\begin{center}
Belle II Collaboration\\
\bigskip
\end{center}

\vspace{\fill}

\begin{abstract}
\noindent We describe the physics potential of the Belle II experiment with electron-positron data corresponding to integrated luminosities of 1\invab to 50\invab. We discuss Belle II's unique capabilities in reconstructing  neutral particles, neutrinos and other ``invisible" particles, and inclusive final states to probe non-standard-model physics. We project sensitivities for compelling measurements that are of primary relevance and where Belle II reach is unique or world leading.
\newpage

\noindent{\bf Editor and corresponding author}\\ 

\noindent \texttt{Diego Tonelli <diego.tonelli@ts.infn.it>}\\

\noindent {\bf Contributors}\\ 

\noindent \texttt{Latika Aggarwal}, 
\texttt{Swagato Banerjee}, 
\texttt{Sunil Bansal}, 
\texttt{Florian Bernlochner}, 
\texttt{Michel Bertemes}, 
\texttt{Vishal Bhardwaj},
\texttt{Alexander Bondar}, 
\texttt{Thomas E.~Browder}, 
\texttt{Lu Cao}, 
\texttt{Marcello Campajola}, 
\texttt{Giulia Casarosa},
\texttt{Claudia Cecchi},
\texttt{Racha Cheaib}, 
\texttt{Giacomo De Pietro}, 
\texttt{Angelo Di Canto}, 
\texttt{Mirco Dorigo}, 
\texttt{Paul Feichtinger}, 
\texttt{Torben Ferber}, 
\texttt{Bryan Fulsom}, 
\texttt{Marcela Garc\'ia},
\texttt{Giovanni Gaudino}, 
\texttt{Alessandro Gaz},
\texttt{Alexander Glazov}, 
\texttt{Svenja Granderath},
\texttt{Enrico Graziani}, 
\texttt{Daniel Greenwald},
\texttt{Pablo Goldenzweig}, 
\texttt{Ivan Heredia},
\texttt{Michel Hern\'andez Villanueva},
\texttt{Takeo Higuchi}, 
\texttt{Thibaud Humair},
\texttt{Toru Iijima}, 
\texttt{Gianluca Inguglia},
\texttt{Akimasa Ishikawa},
\texttt{Daniel Jacobi},
\texttt{Henrik A.~Junkerkalefeld},
\texttt{Robert Karl},
\texttt{Klemens Lautenbach},
\texttt{Peter M.~Lewis},
\texttt{Long-Ke~Li},
\texttt{Stefano Lacaprara},
\texttt{James Libby},
\texttt{Elisa Manoni},
\texttt{Alberto Martini},
\texttt{Mario Merola},
\texttt{Marco Milesi},
\texttt{Stefano Moneta},
\texttt{Minakshi Nayak},
\texttt{Shohei Nishida},
\texttt{Maria Antonietta Palaia},
\texttt{Francis Pham},
\texttt{L\'eonard Polat},
\texttt{Soeren A.~Prell},
\texttt{Elisabetta Prencipe}, 
\texttt{G\'eraldine R\"auber},
\texttt{Isabelle Ripp-Baudot},
\texttt{Markus Röhrken},
\texttt{Michael Roney},
\texttt{Armine Rostomyan},
\texttt{Yoshihide Sakai},
\texttt{Yo Sato},
\texttt{Christoph Schwanda},
\texttt{Alan J.~Schwartz},
\texttt{Justine Serrano},
\texttt{William Sutcliffe},
\texttt{Henrikas Svidras},
\texttt{Kerstin Tackmann},
\texttt{Umberto Tamponi},
\texttt{Francesco Tenchini},
\texttt{Karim Trabelsi},
\texttt{Rahul Tiwary},
\texttt{Kenta Uno},
\texttt{Anselm Vossen}, 
\texttt{Bruce Yabsley}, 
\texttt{Jun-Hao Yin}, and
\texttt{Laura Zani}\\

\vfill

\noindent {\bf Relevant Snowmass topical groups}\\
\begin{description}
\item[RF01] Weak Decays of $b$ and $c$ Quarks
\item[RF02] Weak Decays of Strange and Light Quarks
\item[RF03] Fundamental Physics in Small Experiments
\item[RF04] Baryon and Lepton Number Violating Processes
\item[RF05] Charged Lepton Flavor Violation (electrons, muons and taus)
\item[RF06] Dark Sector Studies at High Intensities
\item[RF07] Hadron Spectroscopy
\item[EF04] EW Physics: EW Precision Physics and Constraining New Physics
\item[EF05] QCD and Strong Interactions: Precision QCD
\item[EF06] QCD and Strong Interactions: Hadronic Structure and Forward QCD
\end{description}

\end{abstract}

\vspace{\fill}

\end{titlepage}

\pagestyle{empty}  


\renewcommand{\thefootnote}{\arabic{footnote}}
\setcounter{footnote}{0}
\tableofcontents
\cleardoublepage
\pagestyle{plain} 
\setcounter{page}{1}
\pagenumbering{arabic}


\graphicspath{{figs/}}
\allowdisplaybreaks

\section{Executive summary}
Belle~II is a particle-physics experiment operating at the intensity frontier. Over the next decades, it will record the decay of billions of bottom mesons, charm hadrons, and $\tau$ leptons produced in 10 GeV electron-positron collisions at the SuperKEKB high-luminosity collider at KEK. These data, collected in low-background and kinematically known conditions, will allow us to measure hundreds of parameters that test the standard model (SM) and probe for the existence of new particles, at mass scales orders of magnitudes higher than those studied at the energy frontier.
We project our sensitivities for measurements that are of primary relevance and where Belle~II will be unique or world leading for data corresponding to 1\invab to 50\invab. \par Belle~II will uniquely probe non-SM contributions in sensitive $b \to q\bar{q}s$ decays and charmless $b \to q\bar{q} d(u)$  decays, semileptonic $b \to s \nu\bar{\nu}~\mbox{and}~s \tau^+\tau^-$ decays, fully leptonic $b \to \ell \nu$ decays, and select $c\to u$ processes. Belle~II will lead exploration of non-SM physics in $b \to c \tau \nu$ and $b\to s\gamma$ decays and will most precisely determine the quark-mixing parameters $|V_{ub}|$ and  $|V_{cb}|$. Belle II will measure many parameters in $\tau$ physics to precisions that will be world leading for the foreseeable future, including the electric and magnetic dipole moments, branching fractions for charged-lepton-flavor-violating decays, and quantities that test  lepton-flavor universality. Belle II will perform unique searches for dark-sector particles with masses in the MeV--GeV range. We will also pursue a broad spectroscopy program for conventional and multiquark  $c \bar{c}$ and $b \bar{b}$ states and provide essential inputs to sharpen the interpretation of muon magnetic-anomaly results. Our exploration of uncharted regions of non-SM parameter space with unprecedented precision will reveal non-SM particles or set stringent constraints on their existence, guiding future endeavors.  \par This document is organized as follows. The introductory sections \ref{sec:landscape}, \ref{sec:detector}, and \ref{sec:methodology} outline the physics landscape, the Belle II detector and its performance, and the methodology adopted for sensitivity projections. Sections \ref{sec:ckm}, \ref{sec:bdecays}, and \ref{sec:charm} discuss precision quark-mixing-matrix tests and non-SM probes based on bottom and charm decays, and are relevant mainly for RF01 and, to a minor extent, for RF04. Section \ref{sec:qcd} discusses measurements associated with quantum-chromodynamics and is relevant for RF03, RF07, EF05 and EF06. Section \ref{sec:tau} is dedicated to $\tau$-lepton physics and is relevant for RF02, RF04 and RF05. Section \ref{sec:dark} deals with direct searches for dark-sector particles and is relevant for RF06. Section \ref{sec:ewk} discusses precision electroweak probes and is relevant for EF04.

\section{Landscape\label{sec:landscape}}
Precision probes have long been essential to advance HEP, due to higher-energy discovery reach compared to {\it direct} searches and capability to characterize the properties of the fields observed in direct searches. High-energy hadron collisions at the LHC have not revealed non-SM physics phenomena at the TeV scale; direct discovery of such phenomena at the energy frontier is unlikely in the near future. In addition, constraints on weakly coupled dark matter at the electroweak scale motivate models with dark-matter particles and mediators with MeV--GeV masses. The Belle II experiment has unique capabilities to look for {\it indirect} phenomena, sensitive to higher energies than direct searches, in heavy quark and $\tau$ physics, and to directly search for dark-matter particles in the MeV to GeV range.

Our predecessors, Belle and \babar, published almost 1200 physics papers, many of which have had seminal impact on our understanding of nature. They significantly furthered our exploration of SM extensions and refined our picture of the weak and strong interactions of quarks. The Belle~II detector is a major improvement over its predecessors. It is located at the SuperKEKB facility at KEK in Tsukuba, Japan, which collides electrons and positrons at  10~GeV with asymmetric energies, and began operations in 2018~\cite{Akai:2018mbz}. Over the next decade, Belle~II will record 50\invab of data, forty times that collected by Belle and \babar. This is made possible by novel low-emittance beams and strong vertical focusing implemented by the new SuperKEKB accelerator, resulting in beam heights of only 60~nm. SuperKEKB is steadily converging to its design goals, as demonstrated by its record peak instantaneous luminosity obtained in late 2021 with relatively moderate currents. Our goal is to record billions of decays of bottom mesons, charm and lighter hadrons, and $\tau$ leptons in kinematically well constrained, low-background conditions. These data enable a program of hundreds of measurements unique and synergic with those from the energy frontier that will extend over a decade and beyond. \par The Belle~II collaboration includes 1100 physicists from 26 countries and regions, including about 80 from 17 US institutions. The US hosts a Tier 1 GRID computing facility at BNL. US groups have played major roles in the design, construction, commissioning, and operations of the hadron- and muon-identification detectors; in critical machine-detector interface studies of beam backgrounds; in management and leadership positions, and in data analysis. The US collaborators are also heavily involved in detector and data-acquisition-system upgrades planned over the next ten years.

\section{The Belle II detector and its performance\label{sec:detector}}

Belle~II is a nearly $4\pi$ magnetic spectrometer~\cite{Belle-II:2010dht} surrounded by a  calorimeter and muon detectors. It comprises several subdetectors arranged cylindrically around the interaction space-point and with a polar structure reflective of the asymmetric distribution of final-state particles resulting from a boosted collision center-of-mass. From innermost out, these subdetectors are the vertex detector, central drift chamber, electromagnetic calorimeter, and $K$-long and muon detector. In between the drift chamber and the calorimeter are charged-particle-identification subdetectors: a time-of-propagation Cherenkov subdetector in the barrel of the cylinder and a aerogel ring-imaging Cherenkov subdetector in the forward region. Between the calorimeter and the muon detector is a solenoid coil that provides a 1.5 T axial magnetic field. The field allows us to measure the momenta and electric charge of charged particles. The vertex detector, built from layers of position-sensitive silicon, samples the charged-particle trajectories (tracks) in the vicinity of the interaction point and allows us to determine the positions of decaying particles. It consists of two layers of pixel sensors surrounded by four layers of microstrip sensors~\cite{Belle-IISVD:2022upf}. 
The second pixel layer is currently incomplete, covering approximately 15\% of the azimuthal acceptance. The observed impact-parameter resolution is typically 10--15~$\mu$m, resulting in 20--30~$\mu$m typical vertex resolution.  The small-cell helium-ethane central drift chamber measures the positions of charged particles at large radii and their energy losses due to ionization. Our relative charged-particle transverse momentum resolution is typically  0.4\%/\pt[GeV/$c$]. The observed hadron identification efficiencies are typically 90\% at 10\% contamination. Typical uncertainties in hadron-identification performance are 1\%. 
The CsI(Tl)-crystal electromagnetic calorimeter measures the energies of electrons and photons with energy-dependent resolutions in the 1.6--4\% range. Layers of plastic scintillators and resistive-plate chambers interspersed between the magnetic flux-return yoke's iron plates allow us to identify $K^0_{\rm L}$ and muons. Our observed lepton-identification performance shows 0.5\% pion contamination at 90\% electron efficiency, and 7\% kaon contamination  at 90\% muon efficiency. Typical uncertainties in lepton-identification performance are 1\%--2\%.\par
Installation of the pixel detector will be completed in 2023. Potential medium- and longer-term Belle~II and SuperKEKB upgrades are under discussion. They will increase the sensitivities of searches for non-SM physics by improving robustness against beam-related backgrounds, providing larger safety factors for design-luminosity operations, increasing subdetector lifetimes, coping with possible future redesigns of the interaction region, and improving overall performance. Details on options and performance are given in a dedicated Snowmass report~\cite{Forti:2022mti}.\par
Belle~II has unique advantages over hadron-collider experiments despite having comparatively less data and fewer accessible initial states~\cite{Belle-II:2018jsg}. SuperKEKB collisions produce $B$ meson pairs with no additional particles. Our backgrounds are typically smaller due to the low multiplicity of final-state particles and the absence of event pile-up. Our reconstruction efficiencies are largely uniform in decay time and kinematic properties of the final states. We reconstruct neutral particles (photon, $\pi^0$, $K^0_S$, $K^0_L$) nearly as well as charged particles. Because the initial state is known and the detector is nearly hermetic, we can reconstruct fully-inclusive final states and broadly search for particles with little or no direct signature in the detector, irrespective of their lifetimes.\par
Vital to our analyses are two high-level algorithms. One identifies the flavor of a neutral $B$ meson at the time of its decay~\cite{ReftoFlavortaggingpaper}. The other reconstruct the kinematic properties of $B$ mesons that are not fully reconstructable from their decay products~\cite{Keck:2018lcd}, using instead information from the rest of the event.   Both reconstruct $B$ decays via thousands of possible decay chains and calculate the quality of the reconstruction. The flavor-tagging algorithm identifies neutral $B$ flavor in 30\% of signal candidates. The rest-of-the-event algorithm finds a $B$ decay 30\%--50\% more frequently than approaches used at previous $B$ factories.\par

\section{Methodology\label{sec:methodology}}
We determine sensitivities for benchmark integrated luminosities (${\cal L}$) of 1~ab$^{-1}$, 5~ab$^{-1}$, 10~ab$^{-1}$, and 50~ab$^{-1}$. 
Whenever possible, we scale existing Belle~II, Belle, and \babar\ results to the baseline 1~ab$^{-1}$ sample size, which is approximately the current combined integrated luminosity of Belle and Belle~II.
We extrapolate future statistical uncertainties by scaling the baseline with ${\cal L}^{-1/2}$ for central confidence intervals and as ${\cal L}^{-1/2}$  to ${\cal L}^{-1}$, depending on expected background conditions, for one-sided intervals.  We classify known systematic uncertainties as luminosity dependent ({\it e.g.}, dependent on the amount of auxiliary data collected) or not and extrapolate them accordingly. Examples of the latter vary broadly across analyses, and include irreducible uncertainties in the description of experimental resolutions, intrinsic limitations associated with Dalitz-plot models of multiparticle kinematics, systematic uncertainties of the corrections needed to match performance determined from simulation with performance observed in data, and others. We attempt at supporting quantitative future projections with observations in data. When realistic and quantitatively supported chances for improving detector and reconstruction performance exist, we include them (labeled baseline, improved, etc). In spite of this cautious approach, the precision of many results remains limited by sample size up to final expected integrated luminosities. For the other results, our projections are conservative, as previous experiments showed that data sets of unprecedented size often offer novel and powerful ways to reduce systematic uncertainties to levels previously unanticipated.\\ Sensitivities obtained from existing Belle II results are derived from event samples reconstructed and analyzed with the Belle II analysis software framework~\cite{basf2, github}.
In what follows, charge-conjugate decays are implied unless otherwise stated; generic particle symbols such as $B$ or $D$ indicate collectively charged and neutral particles; and common shorthands, such as  $K^{*}$ for $K^{*}(892)^{0,+}$, are used to lighten notation when meaning is unambiguous.

\section{Precision CKM tests and searches for non-SM {\it CP} violation in {\it B} decays\label{sec:ckm}}

\subsection{Determination of \texorpdfstring{$\phi_1$}{phi1} from decay-time-dependent {\it CP}-violating asymmetries in {\it B} decays}

Decay-time dependent decay-rate asymmetries offer multiple probes of contributions from massive non-SM particles to the mixing or decay amplitudes. The goal is to test for significant discrepancies between observed asymmetries and asymmetries predicted by Cabibbo-Kobayashi-Maskawa (CKM) hierarchy, or between asymmetries observed in different channels dominated by the same phases in the SM. Using tree-dominated $(c\bar{c})K^0  \equiv J/\psi K_S^0, \psi(2S)K_S^0, \chi_{c1}K_S^0$, and $J/\psi K_L^0$ decays, first generation $B$-factory experiments and LHCb achieved determinations of $\phi_1 \equiv \arg(-V_{cd}V_{cb}^*/V_{td}V_{tb}^*)$ at 2.4\% precision~\cite{HFLAV:2019otj} dominated by systematic uncertainties (typically associated with imperfections in the vertex reconstruction algorithm and flavor tagging~\cite{PhysRevLett.108.171802,PhysRevD.79.072009,PhysRevLett.115.031601}). This offers a reliable and precise SM reference, whose precision is expected to further improve to below 1\% in the next decade. The current goal is to approach that precision in the corresponding channels governed by loop amplitudes to uncover any discrepancies.  Since the most promising loop-dominated channels involve $\pi^0$ or $K^0_S$ in the final states, Belle~II is the premier environment to pursue such measurements. 
Measurements of decay-time-dependent \CP violation are among the most sophisticated in collider physics. They require command of most high-level experimental capabilities including background suppression, vertexing, and identification of the bottom-meson flavor at production. 

\noindent
In the $B^0$-$\overline{B}^0$ system coherently generated from an $\Upsilon(4S)$ decay, 
one $B$ meson (labeled as $B_{\CP}$) may decay to a \CP eigenstate $f_{\CP}$ at $t=t_{\CP}$ whereas
the other (labeled as $B_{\rm tag}$) may decay to a flavor specific final state at $t=t_{\rm tag}$.
The distribution of the proper-time difference $\Delta t\equiv t_{\CP}-t_{\rm tag}$ is expressed by
\begin{equation}
{\cal P}_{f_{\CP}}(\Delta t,q) = \frac{e^{-|\Delta t|/\tau_{B^0}}}{4\tau_{B^0}}\left\{1+q\left[{\cal A}_{f_{\CP}}\cos (\Delta m_d \Delta t) + {\cal S}_{f_{\CP}}\sin(\Delta m_d\Delta t)\right]\right\},
\end{equation}
where $\tau_{B^0}$ and $\Delta m_d$ are the average lifetime and mass difference between neutral $B$ physical states, respectively, and the ${\cal A}_{f_{\CP}}$ and ${\cal S}_{f_{\CP}}$ are the direct and mixing-induced \CP-violating asymmetries, respectively. The $B$ meson flavor $q$  takes values $+1(-1)$ when $B_{\rm tag}$ is $B^0$($\overline{B}^0$) and it is statistically determined from final-state information~\cite{ReftoFlavortaggingpaper}. The time-difference $\Delta t$ is approximated by the distance between the two $B$-meson decay vertices  divided by the speed of the $\Upsilon(4S)$ projected onto the boost axis.\par  
The most promising channel, $B^0 \to \eta'K_S^0$, has a sizable decay rate dominated by the $b\to s$ loop amplitude, where non-SM physics can contribute.  The quantity of interest is $\Delta{\cal S}_{\eta'K_S^0}\equiv{\cal S}_{\eta'K_S^0} -\sin 2\phi_1$, where $\phi_1$ is accurately predicted by CKM hierarchy and measured in tree-amplitude-dominated decays.
SM predictions that include a systematic treatment of low-energy QCD amplitudes assuming factorization yield $0.00<\Delta{\cal S}_{\eta'K_S^0}<0.03$~\cite{Beneke:2005pu}. Establishing a nonzero value of $\Delta{\cal S}_{\eta'K_S^0}$ would indicate non-SM dynamics. The current global value of $\Delta {\cal S}_{\eta'K_S^0}^{\rm exp}$ is $-0.07\pm0.06$~\cite{HFLAV:2019otj}.  Low backgrounds and a high-resolution electromagnetic calorimeter offer Belle~II unique access to this measurement. The precision of the SM prediction, along with the excellent Belle II experimental perspectives,  make $B^0\to\eta'K_S^0$ the most promising channel in this program. 
Similarly promising is the channel $B^0 \to \phi K_S^0$, whose final state makes Belle~II strongly competitive despite challenges associated with model-related systematic uncertainties from the Dalitz plot analysis. In addition, the processes $B^0\to K_S^0\pi^0\gamma$, $B^0\to K^0_S\pi^+\pi^-\gamma$,  and $B^0 \to \rho^0\gamma$ are greatly sensitive to non-SM physics through $b\to s\gamma$ and $b\to d\gamma$ loops and offer Belle~II further exclusive opportunities. \par
\begin{figure}[!ht]
    \centering
    \hspace*{-10mm}
    \includegraphics[width=0.45\textwidth]{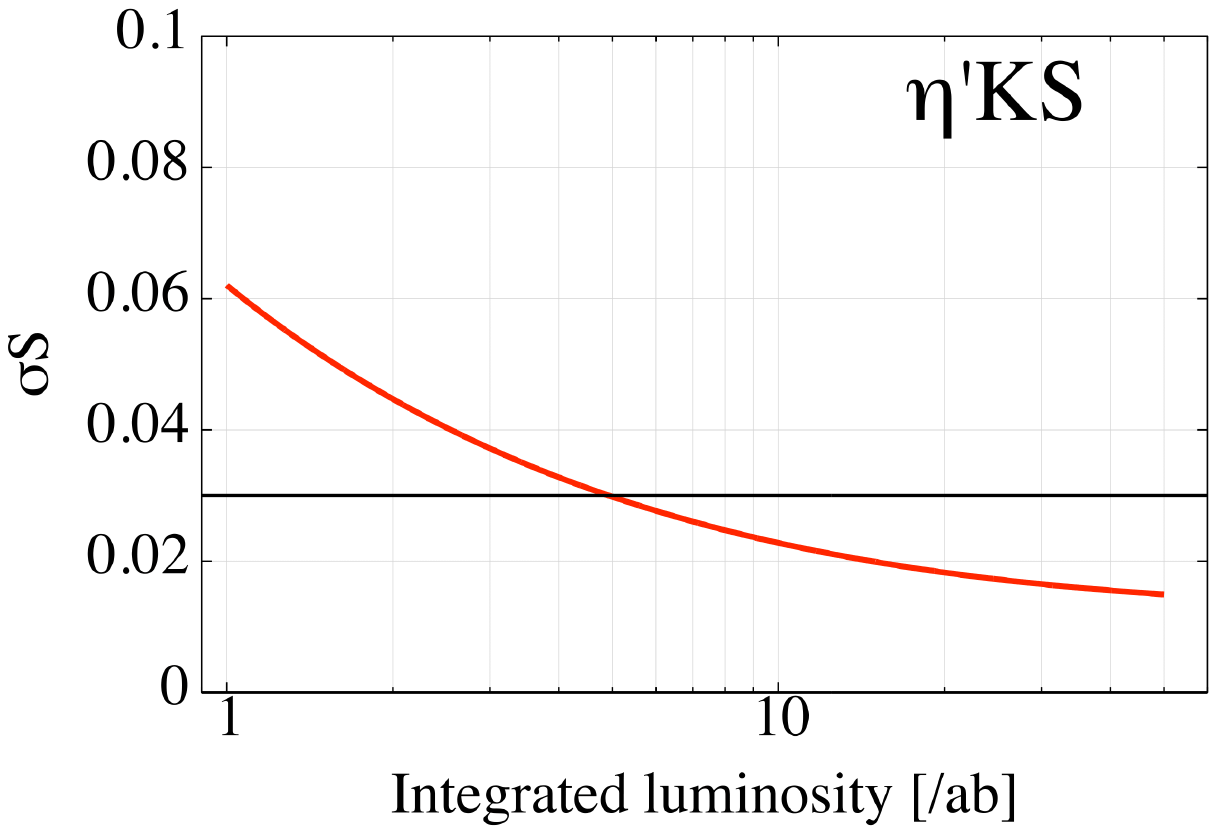}
    \hspace*{3mm}
    \includegraphics[width=0.45\textwidth]{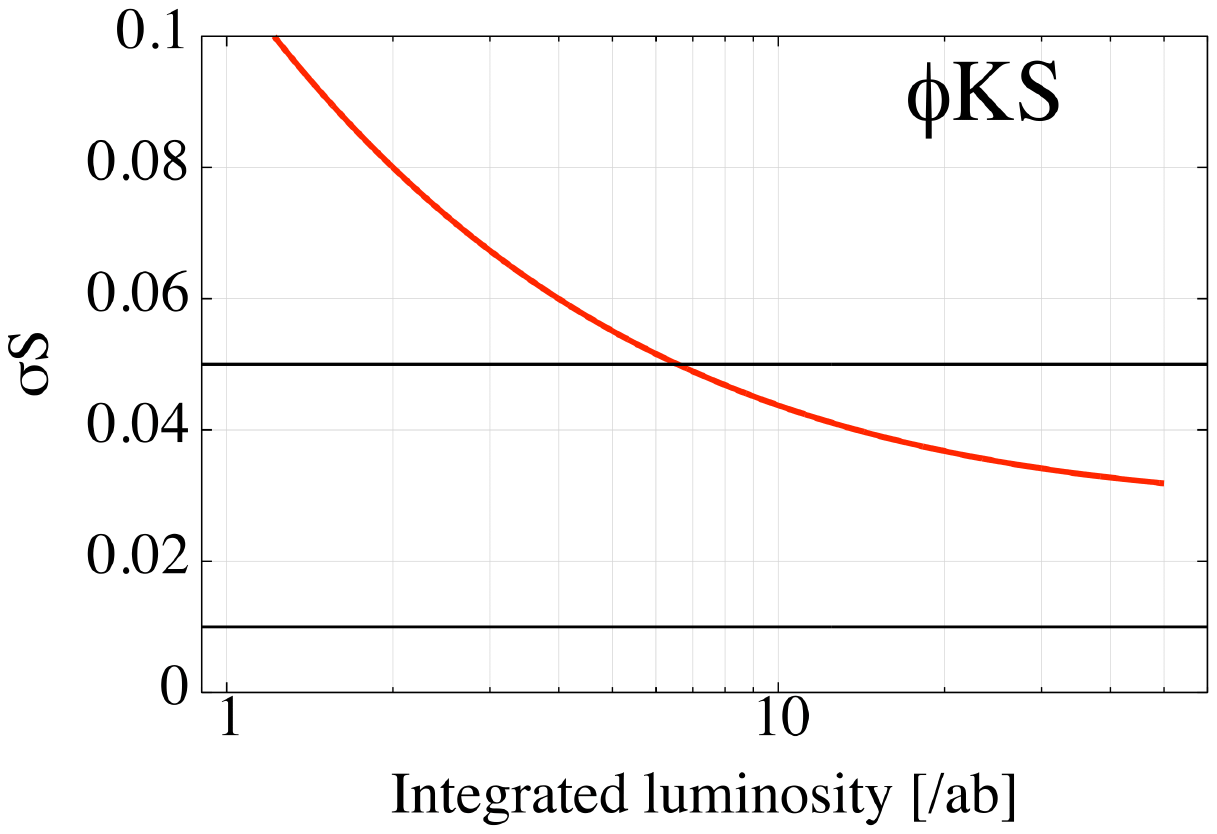}
    \caption{
    Projections of (solid red) total uncertainty in the relevant \CP-violating parameters from (left panel) $B^0\to\eta'K_S^0$  and (right panel) $B^0 \to \phi K_S^0$  decays as a function of the integrated luminosity. Solid horizontal black lines indicate the predicted range of $\Delta{\cal S}_{f_{\CP}}$ based on the SM assumption, $0.00<\Delta{\cal S}_{\eta'K_S^0}<0.03$ and $0.01<\Delta{\cal S}_{\phi K_S^0}<0.05$~\cite{Beneke:2005pu}.
    }
    \label{fig:tdcpv_errors}
\end{figure}

 Projections for asymmetry uncertainties in representative loop-dominated channels are in Table~\ref{tab:TDCPV_errors} and Fig.~\ref{fig:tdcpv_errors}. Uncertainties for tree-dominated $(c\bar{c})K^0$ are also shown as useful common references for ultimate target precision. The statistical uncertainties are 
extrapolated from newly simulated results ($K_S^0\pi^0\gamma$), previously simulated results ($\eta'K_S^0$ and $\rho^0\gamma$~\cite{Belle-II:2018jsg}), or existing Belle results ($\phi K_S^0$~\cite{Belle:2010wis} and $(c\bar{c})K_S^0$~\cite{PhysRevLett.108.171802}).
Systematic uncertainties are projected based on dominant sources in existing results, such as vertex-detector alignment, resolution-function modeling, and extrapolating to future data sets by scaling according to control-sample size and including  conservative improvements on the irreducible sources.
\begin{table}[!ht]
    \centering
    \caption{Summary of projections for statistical and systematic uncertainties of the \CP-violating asymmetries for representative loop-dominated channels.
    In $\phi K^0_S$ final states, \CP-violating asymmetries are expressed in terms of the ``effective" angle $\phi_1^{\rm eff}$ through a decay-time-dependent Dalitz analysis of the $K^+K^-K^0_S$ final state. The uncertainty in the \CP-violating parameter ${\cal S}_{\phi K_S^0}$ is derived from the $\phi_1^{\rm eff}$ uncertainty by assuming ${\cal S}_{\phi K_S^0}= \sin2\phi_1^{\rm eff}$.
    }
    \label{tab:TDCPV_errors}
    \begin{tabular}{ccc | ccc}
    \hline\hline
         $\eta'K_S^0$ & $\sigma^{\rm stat}_{\cal S}$ & $\sigma^{\rm syst}_{\cal S}$ & 
        $\phi K_S^0$ & $\sigma^{\rm stat}_{\cal S}$ & $\sigma^{\rm syst}_{\cal S}$ 
         \\
        \hline
        $1~{\rm ab}^{-1}$  & $0.054$ & $0.031$ &
        $1~{\rm ab}^{-1}$ & $0.103$ &  $0.038$ \\
        $5~{\rm ab}^{-1}$  & $0.024$ & $0.017$ &
        $5~{\rm ab}^{-1}$ & $0.046$ &  $0.030$ \\
        $10~{\rm ab}^{-1}$ & $0.017$ & $0.015$ &
        $10~{\rm ab}^{-1}$ & $0.033$ &  $0.029$ \\
        $50~{\rm ab}^{-1}$ & $0.007$ & $0.013$ & 
        $50~{\rm ab}^{-1}$ & $0.015$ &  $0.028$ \\
    \hline
    \hline
    
         $K_S^0\pi^0\gamma$ & $\sigma^{\rm stat}_{\cal S}$ & $\sigma^{\rm syst}_{\cal S}$ & $\rho^0\gamma$ & $\sigma^{\rm stat}_{\cal S}$ & $\sigma^{\rm syst}_{\cal S}$  \\
        \hline
        $1~{\rm ab}^{-1}$  & $0.31$ & $0.02$ & $1~{\rm ab}^{-1}$ & $0.469$ & $0.133$ \\
        $5~{\rm ab}^{-1}$  & $0.14$ & $0.01$ & $5~{\rm ab}^{-1}$ & $0.210$ & $0.061$ \\
        $10~{\rm ab}^{-1}$ & $0.10$ & $0.01$ & $10~{\rm ab}^{-1}$& $0.148$ & $0.044$  \\
        $50~{\rm ab}^{-1}$ & $0.04$ & $0.01$ & $50~{\rm ab}^{-1}$& $0.066$ & $0.024$ \\
    \hline
    \hline
    
         $(c\bar{c})K^0$ & $\sigma^{\rm stat}_{\cal S}$ & $\sigma^{\rm syst}_{\cal S}$ \\
        \cline{1-3}
        $1~{\rm ab}^{-1}$ & $0.018$ & $0.011$ \\
        $5~{\rm ab}^{-1}$ & $0.008$ & $0.008$ \\
        $10~{\rm ab}^{-1}$& $0.006$ & $0.008$ \\
        $50~{\rm ab}^{-1}$& $0.003$ & $0.007$ \\
    \cline{1-3}\cline{1-3}
    \end{tabular}

\end{table}

\noindent
{\bf Belle~II is the only experiment capable of pursuing these measurements, offering exclusive impact in key parameters sensitive to a broad class of generic SM extensions.} For example, $\Delta{\cal S}_{\eta'K_S^0}$ will be ultimately determined with precision comparable with that of theoretical predictions.
While some of the projections based on today's knowledge seem asymptotically limited by systematic uncertainties such as 
uncertainties in vertex resolution models, many of these are common to loop and $(c\bar{c})K_S^0$ modes and therefore are partially canceled in the differences $\Delta{\cal S}$.

\subsection{Determination of  \texorpdfstring{$\phi_2$}{phi2} and charmless hadronic {\it B} decays}
Studies of charmless $B$ decays give access to $\phi_2 \equiv {\rm arg}[-V_{tb}^{*}V_{td}/V_{ub}^{*}V_{ud}]$, the least known angle of the CKM unitarity triangle, and probe contributions of non-SM dynamics in processes mediated by loop decay-amplitudes. However, unambiguous interpretation of measurements from these decays is spoiled by hadronic uncertainties, which are hardly tractable in perturbative calculations.  Appropriate combinations of measurements from 
decays related by flavor symmetries reduce the impact of such unknowns. Especially fruitful are combinations of measurements from decays related by isospin symmetry, which yield robust direct determinations of $\phi_2$~\cite{Charles:2017evz}. Isospin symmetry also provides so-called sum rules, which are linear combinations of branching fraction and \CP-violating asymmetries that offer null tests of the standard model with precision generally better than 1\%~\cite{Gronau:1990ka,Lipkin:1991st,Gronau:1998ep,Gronau:2005kz}.  Owing to comparably strong reconstruction efficiencies and resolutions of neutral and charged particles, {\bf\boldmath Belle~II has the unique capability of studying jointly, and within the same experimental environment, all relevant final states of isospin-related charmless decays. This places Belle~II in a unique position for determining $\phi_2$ and for testing the SM through isospin sum-rules at unprecedented precision}.
The most promising determination of $\phi_2$ relies on the combined analysis of the decays $B^+ \to  \rho^+\rho^0$, $B^0 \to\rho^+\rho^-$, $B^0\to \rho^0\rho^0$, and corresponding decays into pions. Current global precision of 4 degrees is dominated by $B \to \rho\rho$ data~\cite{HFLAV:2019otj}.  Leveraging efficient reconstruction of low-energy $\pi^0$, improved measurements in $B^+\to \rho^+\rho^0$ and $B^0 \to \rho^+\rho^-$ decays will be unique to Belle~II.  \par We benchmark the expected Belle~II performance using a recent $B^+\to \rho^+\rho^0$ analysis that demonstrates Belle~II performance on par with world-best results~\cite{Humair_moriond2022}.  Systematic uncertainties are dominated by data-simulation mismodeling in angular distributions and $\pi^0$ reconstruction efficiency, which improve with luminosity. Contributions from model assumptions on poorly-known decays involving $a_1$, $f_0$, and non-resonant $\pi\pi$ final states will be mitigated by amplitude analyses. However, these might eventually suffer from irreducible Dalitz-model uncertainties that are hard to estimate to date. Assuming that these contributions will be modest, the combined analyses of $B\to\rho\rho$ decays will lead to a $\phi_2$ precision of about 2.5 degrees in a sample corresponding to 10 ab$^{-1}$. \par
\begin{figure}[t]
 \centering
 \includegraphics[width=1.\textwidth]{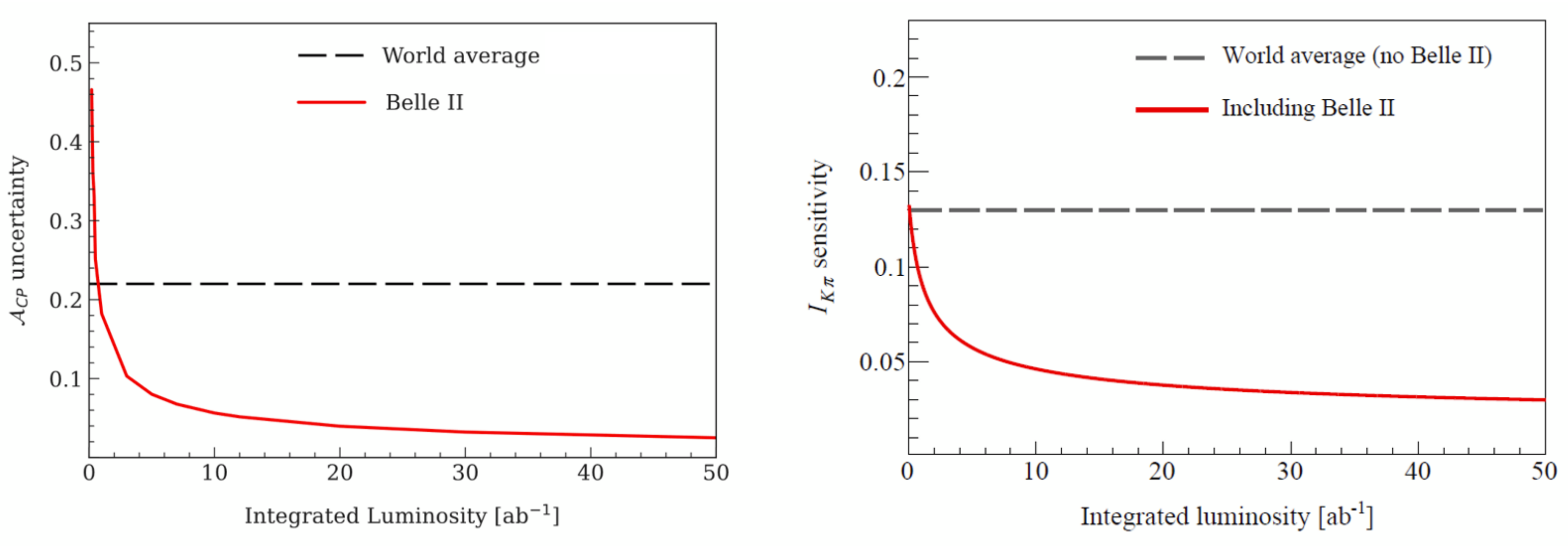}
 \caption{Projected uncertainty as a function of the expected Belle~II sample size on (left panel) the decay-time-integrated \CP-violating asymmetry of $B^0\to\pi^0\pi^0$ decays and (right panel)
 on $I_{K\pi}$ (see text). The inputs for $I_{K\pi}$ are averages of the updates expected from the LHCb and Belle~II experiments combined with current world averages~\cite{PDG}. The solid red curve shows the projection assuming updates on the complete set of $K\pi$ measurements. The dashed grey curve represents the projection assuming no Belle II inputs~\cite{Belle-II:2021jvj}.}
 \label{fig:charmless}
\end{figure}

 Supplementary $\phi_2$ determinations based on $B \to \pi\pi$ decays are limited by the poor knowledge of the branching fraction and direct \CP-violating asymmetry of the $B^0 \to \pi^0\pi^0$ decay. Belle~II is the only experiment to competitively study this channel. Preliminary Belle~II results~\cite{Belle-II:2021dcm} show performance comparable with the world-best results. Since the only irreducible systematic uncertainty is associated with the (sub-percent) precision of branching fractions of control channels used to determine $\pi^0$ reconstruction efficiency on data, this measurement has a ten-fold margin of improvement when extrapolated to the Belle~II data set (Fig.~\ref{fig:charmless}, left). 
An auxiliary measurement of the decay-time-dependent \CP-violating asymmetry in $B^0 \to \pi^0\pi^0$ decays would bring additional insight into $\phi_2$ by suppressing the mirror solutions in the isospin analysis. While restricting to diphoton decays of the $\pi^0$ makes a time-dependent analysis impossible, the full Belle~II data set of 50\,ab$^{-1}$ will enable exploiting photon conversions and $\pi^0$ Dalitz decays~\cite{Belle-II:2018jsg}. Combining all information from $B\to\rho\rho$ and $B\to\pi\pi$ decays (including a decay-time-dependent analysis of $B^0 \to \pi^0\pi^0$) the ultimate uncertainty on $\phi_2$ would be 0.6~degrees with 50\,ab$^{-1}$ of integrated luminosity. This will require a dedicated investigation of isospin-breaking effects~\cite{Falk:2003uq,Charles:2017evz}.\par  Belle~II will also pursue unique determinations of $\phi_2$ based on $B \to \rho \pi$ decays even  though quantitative sensitivity estimates are premature due to the peculiar statistical difficulties associated with multiple, broad likelihood maxima in previous analyses~\cite{PhysRevLett.98.221602, PhysRevD.88.012003}. Such difficulties are not intrinsic of the method to determine $\phi_2$, but result from the size of the samples analyzed at Belle and \babar. 
This strongly motivates an analysis of these decays with a larger data set at Belle II.\par

Equally importantly, Belle~II will provide essential information to address the so-called $K\pi$ puzzle, a long-standing three-standard-deviations anomaly associated with the difference between direct \CP-violating asymmetries observed in $B^0 \to K^+\pi^-$ and $B^+ \to K^+ \pi^0$ decays. 
Belle~II will be unique in measuring the \CP-violating asymmetry in $B^0 \to K^0\pi^0$ decays, the input that  limits the precision of the $I_{K\pi}$ isospin sum rule~\cite{Gronau:2005kz}. This is a dynamical relation among $B$ decay-rates into $K^+\pi^-$, $K^0\pi^+$, $K^+\pi^0$ and $K^0\pi^0$ final states that properly accounts for subleading amplitudes. Since $I_{K\pi} \approx 0$ with $\mathcal{O}(0.01)$ uncertainty in the SM, a precise determination of all inputs offers a reliable and precise null test of the SM. Figure~\ref{fig:charmless} shows a projection of $I_{K\pi}$ sensitivity with and without a Belle~II contribution. 
For measurements of direct \CP asymmetries of $K^+\pi^-$, $K^0\pi^+$, $K^+\pi^0$, we average previous Belle and \babar\  results with projected inputs from LHCb scaled with expected sample size. The $I_{K\pi}$ sensitivity is completely driven by $K^0\pi^0$ inputs, which makes Belle~II essential. Preliminary results from a $B^0 \to K^0\pi^0$ analysis on early data~\cite{Humair_moriond2022} demonstrate Belle~II performance on par with the best Belle result, supporting confidence on future $I_{K\pi}$ sensitivity. Furthermore, analogous to the $K\pi$ system, isospin-sum-rule tests in the $K^\ast \pi$, $K\rho$, and $K^\ast \rho$ systems will provide distinctive non-SM constraints up the largest sample sizes~\cite{Belle-II:2018jsg}.

\subsection{Determination of \texorpdfstring{$\phi_3$}{phi3}}

The phase $\phi_3\equiv\arg[-V_{ud}V_{ub}^*/V_{cd}V_{cb}^*]$ is the only CKM angle accessible via tree-level decays, such as $B\to DK$. Here, $D$ represents a generic superposition of $D^0$ and $\overline{D}^0$. Assuming that non-SM amplitudes do not affect appreciably tree-level processes, precise measurements of $\phi_3$ set strong constraints on the SM description of \CP violation, to be compared  with measurements from higher-order processes potentially sensitive to non-SM amplitudes, such as mixing-induced \CP violation through $\sin(2\phi_1)$. Extraction of $\phi_3$ involves measurement of $B^- \to \overline{D}^{0}K^-$ and  $B^- \to D^{0}K^-$  amplitudes, which are expressed as
\begin{eqnarray*}
\frac{\mathcal{A}(B^-\to \overline{D}^{0}K^-)}{\mathcal{A}(B^-\to D^{0}K^-)} & = & r_{B}e^{i(\delta_B - \phi_3)},\
\label{eq:ratio}
\end{eqnarray*}
where $r_B \approx 0.1$ is the ratio of amplitude magnitudes and $\delta_B$ is the strong-phase difference. Since the hadronic parameters $r_B$ and $\delta_B$ can be determined from data together with $\phi_3$, these measurements are essentially free of theoretical uncertainties~\cite{Brod:2013sga}. The precision of $\phi_3$ is mostly limited by the small branching fractions of the decays involved (around $10^{-7}$). The current global direct determination is  $(66.2^{+3.4}_{-3.6})^{\circ}$~\cite{HFLAV:2019otj}, whereas the indirect determination is $(63.4 \pm 0.9)^{\circ}$~\cite{King:2019rvk}. Any improvement in the direct determination is instrumental in approaching the indirect precision, thus tightening the associated non-SM constraints.  Various methods dependent on the choice of final states accessible to both $D^0$ and $\overline{D}^0$ are used to extract $\phi_3$. 

\begin{figure}[!ht]
\centering
\includegraphics[width=0.8\linewidth]{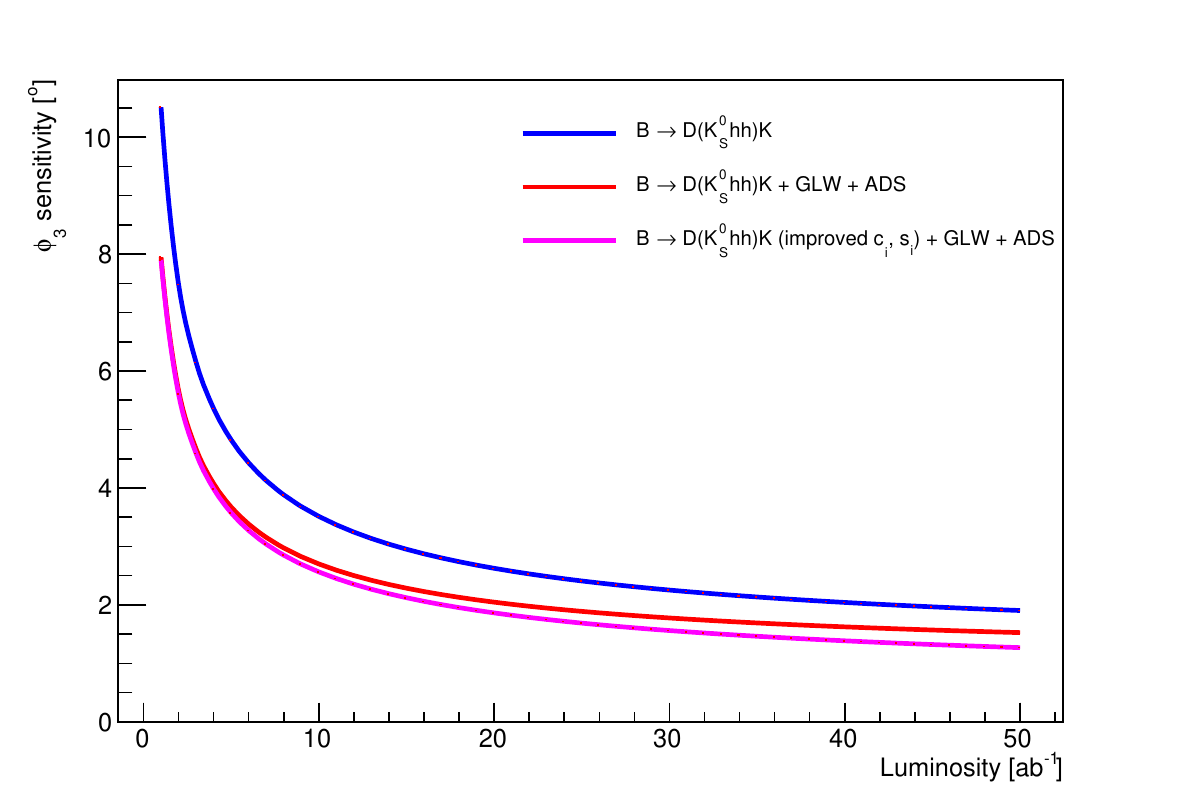}
\caption{Projected uncertainty of $\phi_3$ as a function of expected Belle~II luminosity for three different analysis scenarios.}
\label{fig:phi3_projection} 
\end{figure}

Precision is dominated by measurements based on $B^-\to D (K^{0}_{S}\pi^+\pi^-)K^-$ decays \cite{Giri:2003ty, Bondar:2002da, Belle:2004bbr}. Belle~II will be competitive in this mode and others involving  final-state $K^0_S$, $\pi^0$, and $\gamma$ such as $K^{0}_{S}\pi^0$, $K^{0}_{S}\pi^{+}\pi^{-}\pi^0$ or $B^-\to D^{*}(D{(\gamma, \pi^0})) h^-$. Figure~\ref{fig:phi3_projection} shows projections of $\phi_3$ sensitivity as a function of expected Belle~II luminosity using the  $B^-\to D (K^{0}_{S}h^+h^-)K^-$ channel and its combination with modes including $D$ decays to \CP-eigenstates (known as GLW channels)   \cite{Gronau:1990ra, Gronau:1991dp} and $D^0\to K^{\pm}+n\pi$ final states (known as ADS channels) \cite{Atwood:2000ck}. We extrapolate based on the recent, and current best $B$-factory, measurement of $\phi_3$ for the $B^-\to D (K^{0}_{S}h^+h^-)K^-$ sensitivity~\cite{Belle:2021efh}, and previous Belle measurements for GLW and ADS channels \cite{Trabelsi:2013uj, Belle:2011ac}. 
Extrapolating the current combined Belle and Belle~II measurement allows a straightforward projection of future sensitivity restricted to the $B^-\to D (K^{0}_{S}h^+h^-)K^-$ channel; using this channel only, we expect to achieve a $\phi_3$ sensitivity of $3.5^{\circ}$ and $1.9^{\circ}$ corresponding to Belle~II luminosities of $10 ~\rm ab^{-1}$ and $50 ~\rm ab^{-1}$, respectively. Precision will further improve following the expected three-fold improvements on the external charm-strong-phase inputs from  BESIII~\cite{Belle-II:2018jsg}. In addition,  $B^-\to D (K^{0}_{S}\pi^+\pi^-\pi^0)K^-$ is promising at Belle~II due to its sizable branching fraction and rich resonance substructures, as shown by  Belle~\cite{Belle:2019uav}. Improved charm-strong-phase inputs, availability of a suitable amplitude model of $D \to K^{0}_{S}\pi^+\pi^-\pi^0$ and a larger $B$ decay sample will render $B^-\to D (K^{0}_{S}\pi^+\pi^-\pi^0)K^-$ a strong contributor for determination of $\phi_3$. 

The global Belle~II  precision on $\phi_3$ is foreseen to be $\mathcal{O}(1^{\circ})$ with the full data set. This is comparable to the anticipated precision from LHCb~\cite{LHCb:2018roe}. Similar precision from experiments affected by largely different systematic uncertainties offers complementarity and redundancy; these attributes are crucial when establishing the value of a fundamental parameter that has deep implications on our understanding of SM \CP violation. Improvements in Belle~II detector performance and inclusion of additional, yet-to-be explored modes will likely improve upon this baseline sensitivity.

\newpage

\subsection{Determination of \texorpdfstring{$|V_{ub}|$}{Vub} and \texorpdfstring{$|V_{cb}|$}{Vcb}} \label{sec:VcbVub}

The magnitudes of the CKM matrix elements $|V_{cb}|$ and $|V_{ub}|$ offer stringent constraints in global fits of weak quark interactions, providing reliable SM references to gauge non-SM contributions. The highest-precision determinations of $|V_{cb}|$ and $|V_{ub}|$ come from measurements of rates of semileptonic transitions $b \rightarrow c l \nu$ and $b \rightarrow u l \nu$, either in inclusive or specific final states (exclusive), combined with phenomenological inputs. \par Results from exclusive and inclusive determinations disagree significantly~\cite{HFLAV:2019otj}. 
The reason for this discrepancy is unknown. Most indications point to possibly inconsistent experimental or theory inputs, but also interpretations in terms of non-SM physics cannot be excluded~\cite{Gambino:2020jvv}. The large data set at Belle~II will offer significantly richer, and more precise, experimental information than available thus far 
to overcome the impasse. {\bf Belle II will drive the global $|V_{ub}|$ and $|V_{cb}|$ progress throughout its lifetime.}

\subsubsection{Exclusive \texorpdfstring{$|V_{ub}|$}{Vub}}
Belle~II will help clarifying the experimental status of exclusive $|V_{ub}|$ determinations in a number of ways: while $\overline{B}^0\to\pi^+\ell^-\bar\nu_\ell$ is currently the most effective for determining $|V_{ub}|$ exclusively, Belle~II will measure also other exclusive $b\to u\ell\nu_\ell$ modes with good precision, in particular those involving neutral final-state particles such as $B^-\to(\pi^0,\rho^0,\omega,\eta,\eta')\ell^-\nu_\ell$ and $\overline{B}^0\to\rho^+\ell^-\nu_\ell$.
This will base the exclusive determination of $|V_{ub}|$ on multiple decay channels, thus mitigating experimental and theoretical issues possibly related to $B\to\pi\ell\nu$ only. 
The excellent resolution in $q^2\equiv (p_\ell+p_\nu)^2$ compared to hadron collider experiments (better than $0.45~\gev^2/c^2$ in the worst case for $\overline{B}^0\to\pi^+\ell^-\bar\nu_\ell$) gives access to the decay form factors that are equally important for determining  $|V_{ub}|$. Finally, Belle~II will measure absolute branching fractions and can thus determine $|V_{ub}|$ directly, in addition to the $|V_{ub}|/|V_{cb}|$ ratio.
Current best constraints on $|V_{ub}|$ are reported by $B$-factory experiments in analyses where the (non-signal) partner $B$-meson is either reconstructed ~\cite{Belle:2013hlo} or not~\cite{PhysRevD.83.032007}. Typically, experimental uncertainties are smallest for low $q^2$ whereas uncertainties in the form factors from lattice QCD are smallest at high $q^2$. Improvements in the experimental constraints will be driven largely by data set sizes. \par To illustrate the power of the Belle~II data set, we study the decay $\overline{B}^0\to\pi^+\ell^-\bar{\nu_{\ell}}$. Analyses rely on the reconstruction of the final-state pion and charged lepton with missing four-momentum consistent with a single neutrino. In analyses where the partner $B$ meson is reconstructed (tagged), the signal $B$ rest-frame is determined by the partner $B$ information. Otherwise, it is inferred through kinematic approximations (untagged). The branching fraction differential in $q^2$ is measured to determine the form-factor parameters. \par
\begin{figure}[!ht]
\centering
\includegraphics[width=.5\linewidth]{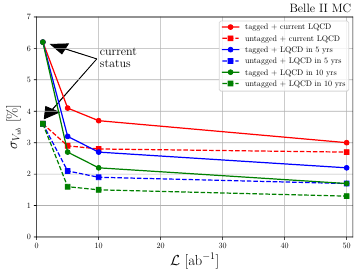}
\includegraphics[width=.37\linewidth]{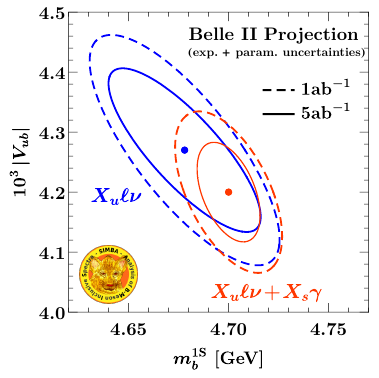}
\caption{(Left panel) projections of $|V_{ub}|$ uncertainties as functions of integrated luminosity in $\overline{B}^0\to\pi^+\ell^-\bar{\nu_{\ell}}$ analyses where the partner $B$ meson is reconstructed (tagged) and not reconstructed (untagged) for current and future expected lattice QCD  inputs~\cite{Belle-II:2018jsg}; (right panel) global fit projections for inclusive $|V_{ub}|$ for 1 ab${}^{-1}$ and 5 ab${}^{-1}$~\cite{Belle-II:2018jsg}. Theoretical uncertainties are not included in the fit.
}
\label{fig:Vub}
\end{figure}

 
Figure~\ref{fig:Vub} (left panel) shows the projections. In all cases, fractional uncertainties of 3\% or better are expected. The precision of tagged determinations is ultimately limited by the precision of the calibration of the partner $B$ reconstruction efficiency. {\bf Belle~II will double the global  precision in exclusive $|V_{ub}|$ results} even if there will be no improvement in theoretical inputs. Expected progress in lattice QCD~\cite{Boyle:2022uba,Belle-II:2018jsg} will offer further significant improvement.

\subsubsection{Inclusive \texorpdfstring{$|V_{ub}|$}{Vub}}

Measurements of inclusive $B \to X_{u} \ell \nu$ decays, where $X_{u}$ is a charmless hadronic system, 
are unique to $B$ factories. Because of on-threshold $B\overline{B}$ production, after reconstructing a signal lepton and the partner $B$ meson, all remaining tracks and energy clusters can be associated with the $X_{u}$ candidate. Measurements are challenging and require accurate modeling of the $b\to u$~signal and the $b\to c$~background as demonstrated in the latest Belle measurement of $B \to X_{u} \ell \nu$, which indeed reports results closer to exclusive~\cite{Belle:2021eni}. 
With larger sample sizes and continuing developments in reconstruction algorithms ({\it e.g.}, improved partner $B$ reconstruction) {\bf Belle~II will accomplish measurements of inclusive $|V_{ub}|$ to $\mathcal{O}(0.01)$ precision} (Table~\ref{tab:inculVub}).
\begin{table}[!ht]
\renewcommand\arraystretch{1.2}
	\centering
\begin{tabular}{lccccc}
\hline \hline 
   & Statistical & Systematic & Total expt. & Theory & Total \\
    &  &  (reducible, irreducible) &  &  \\
\hline
$1~\mathrm{ab}^{-1}$ & $2.5$ & $(2.9,1.6)$ & $4.1$ & $2.5 - 4.5$ & $4.8-6.1$ \\
$5~\mathrm{ab}^{-1}$ & $1.1$ & $(1.3,1.6)$ & $2.3$ & $2.5 - 4.5$ & $3.4 - 5.1$ \\
$50~\mathrm{ab}^{-1}$ & $0.4$ & $(0.4,1.6)$ & $1.7$ & $2.5 - 4.5$ & $3.0 - 4.8$ \\
\hline\hline
\end{tabular}
	\caption{\label{tab:inculVub} Expected fractional uncertainties (in percent) for inclusive $|V_{ub}|$ measurements \cite{Belle-II:2018jsg}.}
\end{table}

Promising novel ideas are also explored. Belle has recently demonstrated a first measurement of the $B \to X_{u} \ell \nu$ differential spectra~\cite{Belle:2021ymg}, paving the way for extensions at Belle~II. These measurements combined with model-independent theory approaches~\cite{Bernlochner:2020jlt, Gambino:2016fdy} allow to determine the leading-order shape-function, needed to extract $|V_{ub}|$ from the partial branching fraction. A key potential here it to use the experimentally most precise regions (lepton endpoint, high $q^2$) alone to determine $|V_{ub}|$ avoiding $b \to c \ell \nu$ systematic uncertainties. 
Figure~\ref{fig:Vub} (right) shows how {\bf Belle~II will further improve inclusive $|V_{ub}|$ determinations} by a global fit using two  one-dimensional differential spectra of hadronic mass and lepton energy for $B \to X_{u} \ell \nu$ and a photon energy spectrum of $B \to X_{s} \gamma$.\par  The tension between determinations of $|V_{ub}|$ based on different methods has eluded the scientific community for a decade. Belle~II is in a unique position to determine the underlying cause. 

\subsubsection{Determination of \texorpdfstring{$|V_{cb}|$}{Vcb}}
While we focus this discussion on the $|V_{ub}|$~program, which stretches longer into the future, 
similar considerations and expected advances apply to determinations of $|V_{cb}|$. The $b \to c$ coupling strength is no less relevant physics-wise~\cite{Charles:2020dfl} and {\bf Belle II has an edge over any existing or foreseen experiment} with the ability to make precise determinations of $|V_{cb}|$ from exclusive $B \rightarrow D^{(*)} \ell \nu$ decays and inclusive $B \rightarrow X_{c} \ell \nu$ decays. 

For exclusive analyses the key experimental challenges will be to understand better the composition and form factors of $B \rightarrow D^{**} \ell \nu$ decays and reduce relevant systematic uncertainties as those associated with lepton identification and low-momentum-pion reconstruction for $B \rightarrow D^{*} \ell \nu$ decays. Belle II will tackle this with a detailed program based on dedicated auxiliary studies of $B \rightarrow D^{**} \ell \nu$ decays.  Precision of inclusive determinations, which is limited by theory, will benefit from measurements of the kinematic moments of $B \rightarrow X_{c} \ell \nu$ decays that will constrain hadronic matrix elements in the operator-product-expansion based theory. {\bf Ultimately Belle~II will accomplish measurements of $|V_{cb}|$ to $\mathcal{O}(0.01)$ precision}. 

\section{Non-SM probes from semileptonic, radiative, and leptonic {\it B} decays\label{sec:bdecays}}

A number of persistent anomalies have been observed in $B$~meson decays with missing energy:  an apparent deviation from lepton-flavor universality in the decays $B\to D^{(*)}\tau\nu_\tau$ consistently stayed at the $3\sigma$ level since these decays were first measured~\cite{HFLAV:2019otj}. {\bf The unique capability of Belle~II to reconstruct final states with missing energy and identify efficiently all species of leptons will considerably improve the understanding of these anomalies.}

\subsection{Semitauonic {\it B} decays}
Decays $B \rightarrow D^{(*)} \tau \nu_{\tau}$ offer precious opportunities for testing lepton-flavor universality at high precision opening a window onto lower-mass (TeV range) non-SM particles. Sensitive observables are the ratio $R(D)$ and $R(D^*)$ between branching fractions of $B \rightarrow D^{(*)} \tau \nu_{\tau}$ and $B\rightarrow D^{(*)} \ell \nu_{\ell}$ decays, where  $\ell=e$ or $\mu$. 
Current best results on $R(D^{(*)})$ are reported by the Belle experiment~\cite{Belle:2019rba} and are consistent with previous measurements~\cite{BaBar:2013mob,LHCb:2015gmp,LHCb:2017rln,Belle:2015qfa,Belle:2017ilt} in showing a (combined) 3.1$\sigma$ excess with respect to the SM expectation~\cite{HFLAV:2019otj}.
This deviation attracted significant interest geared at understanding whether it could be a potential indication of non-SM dynamics or rather generated by mundane effects, as poorly understood feed-down from excited $D$ mesons. 
Investigating the anomaly through precision measurements of $R(D^{(*)})$ is a chief goal of Belle~II. A data set of unprecedented size, combined with detector, reconstruction, and analysis improvements will be instrumental in providing critical insight with {\bf precision superior to any other experiment}. The main experimental challenge is achieving a detailed understanding of poorly known $B\rightarrow D^{**} \ell \nu$ backgrounds, whose feed-down may bias the results. The anticipated data set size will allow for accurate tagged measurements of $B\rightarrow D^{**} \ell \nu$ decays for several $D^{**}$ states using samples reconstructing on the signal-side a lepton, a $D^{(*)}$ meson and $n$ pions. If a non-SM source of the anomaly would be established, angle-dependent asymmetries and differences between forward-backward asymmetries observed in muons and electrons, which are ideally suited for Belle~II, may offer insight on the properties of the non-SM couplings involved.


Furthermore, Belle~II has the unique potential to also measure the ratio of the inclusive rate $B \rightarrow X \tau \nu$ to the lower-mass lepton counterparts $R(X)$. This probes both electron and muon modes with a precise consistency check whose phenomenological interpretation is independent from lattice-QCD uncertainties that affect the other observables. However, this is a challenging measurement attempted and never completed at previous $B$-factory experiments. The main experimental challenge is to control the significant systematic uncertainties associated with background composition. Auxiliary measurements of $B\rightarrow D^{**} \ell \nu$ and other backgrounds will be essential. Belle~II will also pursue measurements of semitauonic $b\to u$~decays that were not accessible with previous $B$-factory data sets and offer the first probe of the anomaly in $b\to u$ transitions through the $R(\pi)$ observable. \par
\begin{figure}[!ht]
    \centering
    \includegraphics[width=10cm]{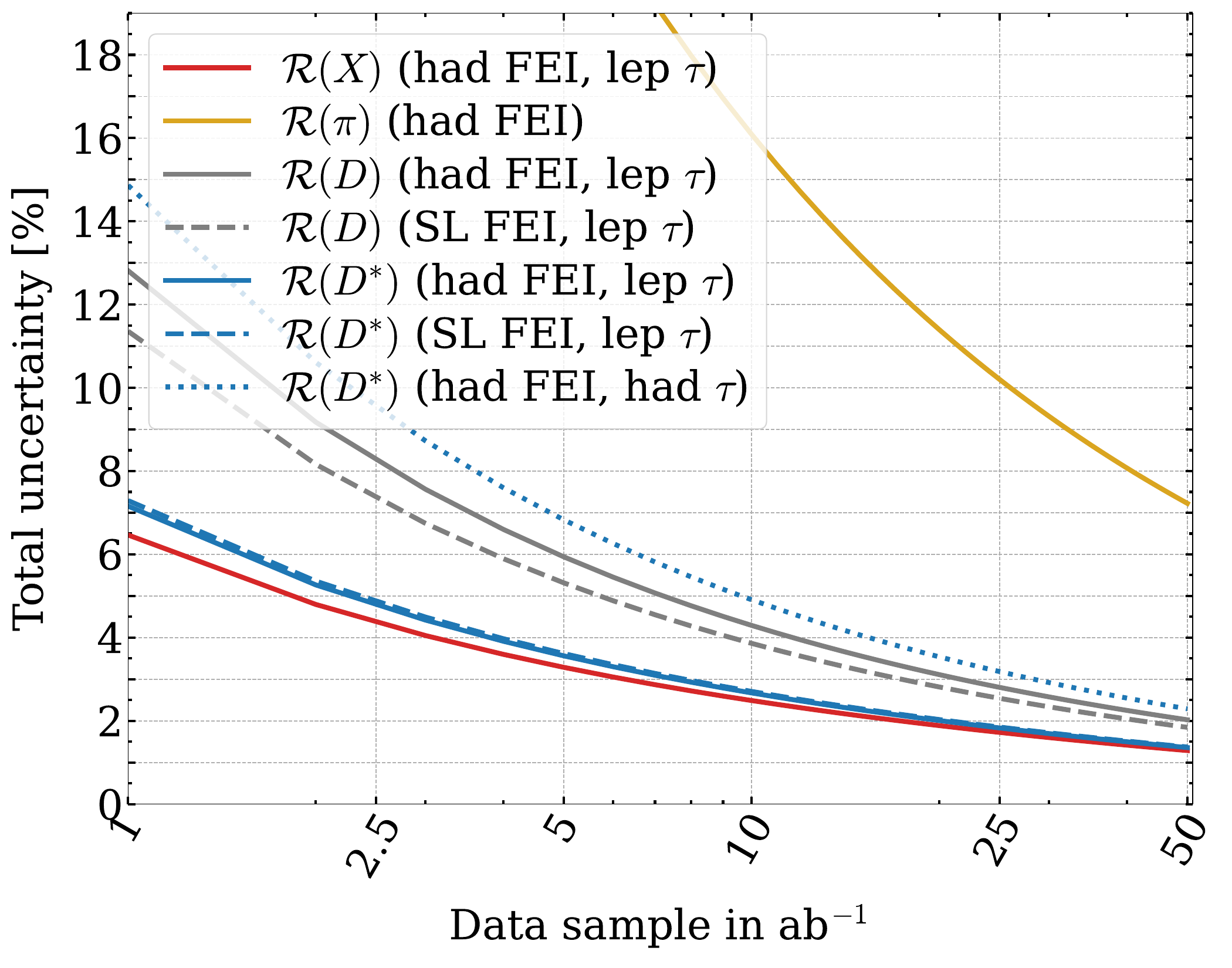}
    \caption{Expected Belle~II sensitivity for various $R$ measurements as a function of luminosity. The FEI acronym refers to the algorithm for reconstruction of the partner $B$-meson mentioned in Sec.\ 3.}
    \label{RDstarProjections}
\end{figure}

 Figure \ref{RDstarProjections} shows the expected Belle~II sensitivities in measurements of relevant ratios of semileptonic branching fractions as a function of luminosity, based on existing Belle and Belle~II studies~\cite{Belle:2019rba,Belle:2015qfa,Belle:2017ilt}. {\bf Belle II will achieve $\mathcal{O}(10^{-2})$ sensitivities on most relevant observables.}

New approaches will also be explored. Measurements of polarization of the $\tau$ lepton ($(\Gamma^+-\Gamma^-)/(\Gamma^+ + \Gamma^-$)) and $D^*$ mesons ($\Gamma_L/(\Gamma_T + \Gamma_L$)) provide supplementary sensitivity to non-SM physics.  Here, $\Gamma^+(\Gamma^-)$ is the semitauonic decay rate where the $\tau$ has $+\frac{1}{2}$ $(-\frac{1}{2})$ helicity and $\Gamma_L(\Gamma_T)$ is the rate where the $D^*$ has longitudinal (transverse) polarization.   Furthermore, differential angular distributions in $B \rightarrow D^{(*)} \tau \nu$, usually studied as functions of $q^2$, may also important to decipher the dynamics and are distinctive to Belle~II. Combined with determination of $R(D^{(*)})$ and $R(X)$, they provide a  leading semitauonic agenda for precision tests for non-SM physics.\par
\begin{figure}[!ht]
\centering
\includegraphics[width=.75\linewidth]{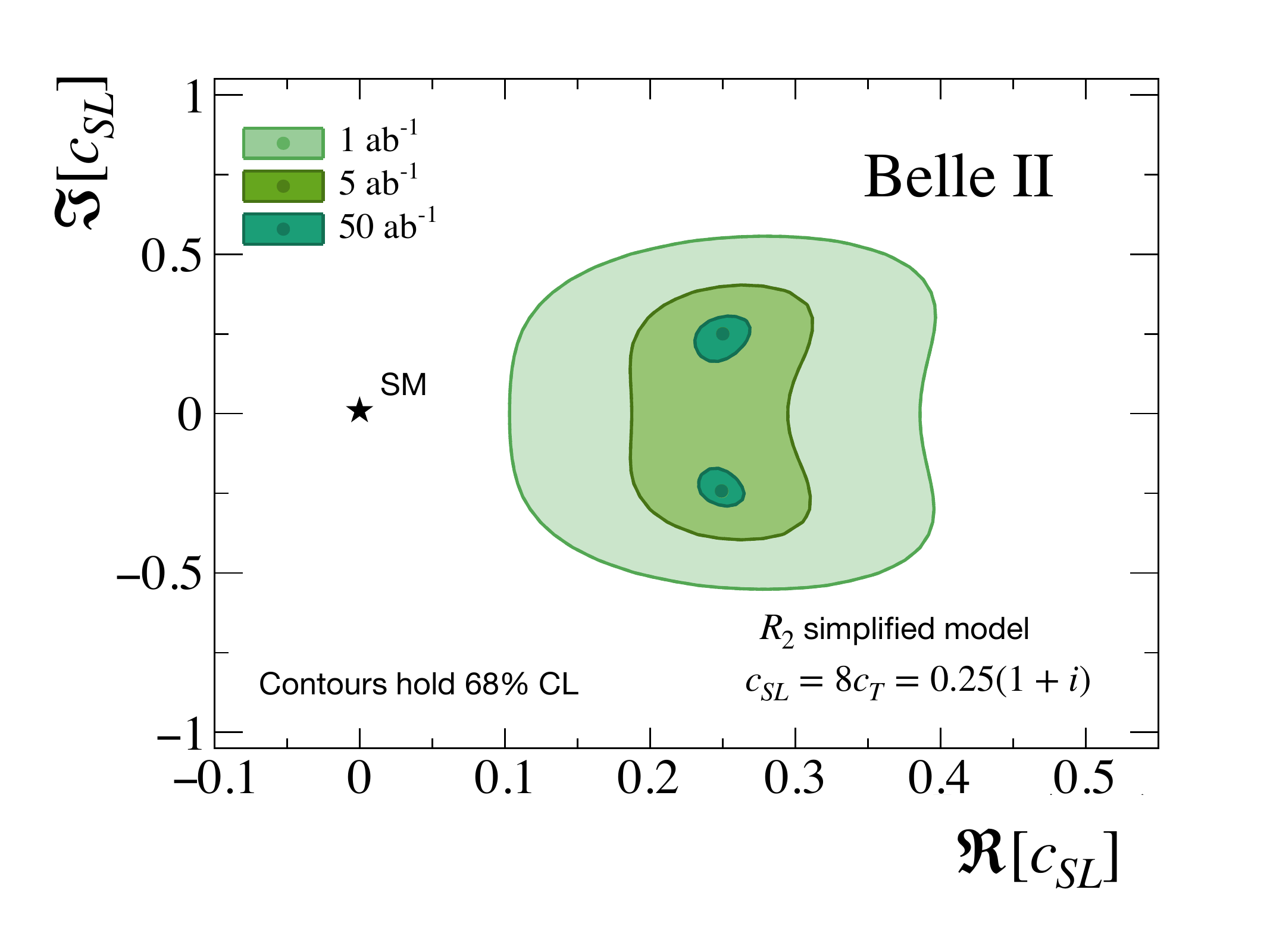}
\caption{Allowed 68\% CL regions for the $R_2$ simplified model coupling $c_{SL} = 8 c_T$~\cite{Freytsis:2015qca,Dorsner:2016wpm} based on fitting to an Asimov data set with $c_{SL} = 0.25(1+i)$ and assuming a Belle~II luminosity ranging from 1 to 50 ab${}^{-1}$~\cite{Bernlochner:2020tfi}. The best fit points are shown as green dots. Assuming $\mathcal{O}(1)$ couplings, this would correspond to a leptoquark of mass around 1.4 TeV.}
\label{fig:wilson}
\end{figure}

 
An illustration of Belle~II's potential is provided by the striking SM discrimination offered by Belle~II data against a simplified 
leptoquark model with coupling $c_{SL} = 8 c_T$~\cite{Freytsis:2015qca,Dorsner:2016wpm}. 
Such models have been proposed recently as economic explanations for the observed enhancements in various $b \to c \tau \bar \nu_\tau$ and $b \to s \ell \ell$ measurements.  Figure~\ref{fig:wilson} shows the anticipated Belle~II sensitivity on  $c_{SL}$~\cite{Bernlochner:2020tfi}. The SM could be excluded with 5~ab${}^{-1}$; 50~ab${}^{-1}$ would allow resolving the coupling up to a two-fold degeneracy. \par 
Regardless of the challenges associated with measuring $R(X)$, {\bf Belle~II will provide the most precise experimental information to resolve the $R(D)$ and $R(D^*)$ anomalies}.


\subsection{\texorpdfstring{$B\to K^{(*)}\nu\bar{\nu}$}{KandKstarnunu} decays} 
\def\BKSnn                      {\ensuremath{B\to K^{(*)}\nu\bar{\nu}}\xspace}
\def\BKpnn                      {\ensuremath{B^{+}\to K^{+}\nu\bar{\nu}}\xspace}
\def\BKznn                      {\ensuremath{B^0\to K^0_{\mathrm S}\nu\bar{\nu}}\xspace}
\def\BKSznn                     {\ensuremath{B^0\to K^{*0}\nu\bar{\nu}}\xspace}
\def\BKSpnn                     {\ensuremath{B^{+}\to K^{*+}\nu\bar{\nu}}\xspace}
The study of flavor-changing neutral-current transitions, such as $b\to s \nu\bar{\nu}$, is a keystone of the Belle~II physics program. These transitions are suppressed in the SM~\cite{Glashow:1970gm} and only occur at higher orders in SM perturbation theory via weak-interaction amplitudes involving the exchange of at least two gauge bosons.
The absence of charged leptons in the final state reduces the theoretical uncertainty compared to $b\to s \ell\ell$ transitions~\cite{Altmannshofer:2009ma,Buras:2014fpa}. 
SM branching fractions range between $2.2\times 10^{-6}$ and $8.4\times 10^{-6}$ depending on final state, with $O(10\%)$ fractional  uncertainties~\cite{Buras:2014fpa}. Decays \BKSnn are of particular interest as they offer complementary probes of non-SM physics scenarios proposed~\cite{Descotes-Genon:2020buf} to explain the anomalies observed in $b\to s\ellell$ transitions~\cite{PhysRevLett.122.191801,Aaij_2017,PhysRevLett.111.191801,Aaij_2014,Aaij_2016,Aaij_2020}.
More generally, \BKSnn decays provide provide discriminating constraints among various non-SM extensions such as models with leptoquarks~\cite{Becirevic:2018afm,Browder:2021hbl},
axions \cite{MartinCamalich:2020dfe},
feebly interacting \cite{Darme:2021qzw},
or dark matter particles~\cite{Filimonova:2019tuy,Alonso-Alvarez:2021qfd}. 


Serious experimental challenges accompany the reliability of theoretical predictions. Final-state neutrinos  leave no signature in the detector and provide no information about the signal \B meson. Indirect information is obtained by reconstructing the (non-signal) partner \B meson produced in the $e^+e^- \to \Y4S \to \B\bar{\B}$. Previous studies explicitly reconstructed the partner \B meson in hadronic~\cite{PhysRevLett.86.2950,PhysRevD.87.111103,PhysRevD.87.112005} or in semileptonic decays~\cite{PhysRevD.82.112002,PhysRevD.96.091101}. Recently, we introduced a novel, \textit{inclusive reconstruction} method~\cite{Belle-II:2021rof} where tracks and energy deposits not associated with the signal candidate are associated with the decay of the accompanying \B meson, or``rest of event". 
\begin{table}[!ht]
  \begin{tabular}{ccccc}
    \hline\hline
     Decay & 1\invab & 5\invab & 10\invab & 50\invab \\
     \hline
      \BKpnn & 0.55 (0.37) & 0.28 (0.19) & 0.21 (0.14) & 0.11 (0.08)\\
      \BKznn & 2.06 (1.37) & 1.31 (0.87) & 1.05 (0.70) & 0.59 (0.40)\\
      \BKSpnn & 2.04 (1.45) & 1.06 (0.75) & 0.83 (0.59) & 0.53 (0.38) \\
      \BKSznn & 1.08 (0.72) & 0.60 (0.40) & 0.49 (0.33) & 0.34 (0.23)\\
     \hline\hline
  \end{tabular}
  \caption{\label{tab:b2knnbase}Baseline (improved) expectations for the uncertainties on the signal strength $\mu$ (relative to the SM strength) for the four decay modes as functions of data set size.}
\end{table}
The inclusive approach yields significantly higher signal efficiency and better sensitivity than any previous approach, as shown by the Belle~II \BKpnn branching fraction results~\cite{Belle-II:2021rof}.\par 
We project sensitivities based on Belle II simulation and an early Belle II analysis~\cite{Belle-II:2021rof}. 
Two scenarios are considered,  which are similar for all except the \BKSpnn decay. The "baseline" scenario assumes no further improvements. The "improved" scenario assumes a 50\% increase in signal efficiency for the same background level, an advance that current studies indicate to be achievable by various means including combination with semileptonic and hadronic reconstruction of the partner $B$ meson. For the \BKSpnn decay, we assume 20\% and 70\% improvements for the baseline and improved scenarios, respectively, since Ref.~\cite{Belle-II:2021rof} relied on the $\Kstarp \to \Kp \piz$ sub-decay only. The projections are in ~\autoref{tab:b2knnbase}. {\bf Belle~II is the only experiment capable of exploring these key channels that disclose a vast and uncharted region of SM and non-SM dynamics.} For example, with just 5\invab of integrated luminosity, the \BKpnn analysis is sensitive to the SM rate at $3\sigma$ ($5\sigma$) level for the baseline (improved) scenarios.

\subsection{\texorpdfstring{$B\to K^{(*)}\ell^+\ell^{(')-}$}{KandKstarellell} decays} 
The transitions $b\to s \mu\mu$ and $b\to s ee$ are under extensive experimental
investigation due to several observed anomalies~\cite{PhysRevLett.122.191801,Aaij_2017,PhysRevLett.111.191801,Aaij_2014,Aaij_2016,Aaij_2020} that prompted interpretations in terms of $O(10)$\tev non-SM particles.  
The unique feature of Belle II is high efficiency and similar performance for muons and electrons, along with access to absolute branching fractions. Based on a recent Belle II analysis~\cite{Manoni_moriond2022}, we expect to provide distinctive information
to assess independently the existence of the anomalies (at current central values) with samples of 5\invab to 10\invab of data.\par
More distinctive is Belle II's reach on $b\rightarrow s \tau\tau$ transitions. These can be enhanced, by up to three orders of magnitude, in several SM extensions that allow for lepton-flavor universality violation in the third generation~\cite{PhysRevLett.120.181802, Alonso2015}. The SM branching fraction for the $B\rightarrow K^{\ast}\tau\tau$ decay is around $10^{-7}$ \cite{PhysRevD.53.4964}, much smaller than current experimental upper limits, which are at around $2.0 \times 10^{-3}$ at 90\% CL \cite{bellecollaboration2021search,PhysRevLett.118.031802}. The presence of two $\tau$ leptons in the final state makes access to these decays ideally suited to Belle~II.\par 
Sensitivity projections are reported for the $B^0 \rightarrow K^{\ast 0} \tau \tau$ decay assuming  reconstruction of the partner $B$ in a fully hadronic 
decay~\cite{Keck:2018lcd} and restricting both $\tau$ leptons to
decay into leptons. Results are then extrapolated to the other choices of partner $B$ reconstruction and decay modes. In the baseline scenario we mirror the Belle analysis~\cite{bellecollaboration2021search}. In an improved scenario where $\tau\rightarrow\pi\nu$ decays are included, signal efficiency is significantly enhanced. In both scenarios we conservatively assume systematic uncertainties from the Belle analysis, which are dominated by limited size of simulated samples and imperfect knowledge of partner $B$ reconstruction efficiency.\par ~\autoref{table:K*0tautau_projections} reports the projections, which show up to a factor of four improvements over current bounds.
\begin{table}[!ht]
	\caption{Projected branching fraction upper limits for the $B^0 \rightarrow K^{\ast 0} \tau \tau$ search in two scenarios (see text).}
	\label{table:K*0tautau_projections}	
	\begin{center}
		\begin{tabular}{l c c}
			\hline\hline
			& \multicolumn{2}{c}{$\mathcal{B}(B^0 \rightarrow K^{\ast 0} \tau \tau) $ (had tag)} \\
			ab$^{-1}$ & "Baseline" scenario  & "Improved" scenario \\
			\hline
			1 & $<3.2 \times 10^{-3}$ & $<1.2 \times 10^{-3}$ \\
			5 & $<2.0 \times 10^{-3}$ & $<6.8 \times 10^{-4}$ \\
			10 & $<1.8 \times 10^{-3}$ & $<6.5 \times 10^{-4}$ \\
			50 & $<1.6 \times 10^{-3}$ & $<5.3 \times 10^{-4}$ \\
			\hline\hline
		\end{tabular}
	\end{center}
\end{table}
A further increase in sensitivity is expected upon inclusion of the charged  $B^+ \rightarrow K^{\ast+} \tau \tau$ channel.  {\bf Belle~II will offer unprecedented sensitivity in $B^+ \rightarrow K^{\ast+} \tau \tau$ decays}. Similar consideration of privileged access and world-leading sensitivity apply to searches for lepton-flavour violating decays such as $B\to X \tau \mu$ and $B \to X \tau e$~\cite{Kumar:2018kmr}.

\subsection{Radiative {\it B} decays}
Radiative $b \to  s \gamma$ transitions are dominated by a one-loop  amplitude involving a $t$ quark and $W$ boson. 
Extensions of the SM predict particles that can contribute to the loop, potentially altering various observables from their SM predictions~\cite{BSM1,BSM2}. 
{\bf Belle~II has a unique capability to study these transitions both inclusively and using specific channels.}

\subsubsection{Inclusive \texorpdfstring{$B\to X_s\gamma$}{Bxsgamma}}
\def\BtoXsdgamma{{\ensuremath{B\rightarrow X_{sd} \gamma}}\xspace}
\def\BtoXsgamma{{\ensuremath{B\rightarrow X_{s} \gamma}}\xspace}
\def\BptoXsgamma{{\ensuremath{B^+\rightarrow X_{s} \gamma}}\xspace}
\def\BztoXsgamma{{\ensuremath{B^0\rightarrow X_{s} \gamma}}\xspace}
\def\btosgamma{{\ensuremath{b\rightarrow s\gamma}}\xspace}
The availability of precise and reliable SM predictions of inclusive \BtoXsgamma rates, where $X_s$ identifies a particle or system of particles with strangeness, make these rates sensitive probes for non-SM physics. In addition, these analyses enable the determination of observables like the $b$-quark mass and can provide input to inclusive determinations of $|V_{ub}|$~\cite{Belle-II:2018jsg}.
Ability to measure precisely the decay properties of the partner $B$ recoiling against the signal $B$ is key for inclusive analyses\cite{Keck:2018lcd}. Current best results show 10\% fractional precision mostly limited by systematic uncertainties associated with understanding the large backgrounds.\par 
In the following projections, the signal $B$ meson is required to decay into a high-energy photon. No constraints are imposed on the hadronic system $X_s$. The partner $B$ meson is reconstructed in its hadronic decays. The lower photon-energy threshold, $E^B_\gamma>1.4~\gev$,  is inferior to those used in previous results, resulting in a significantly more challenging analysis. Lower thresholds accept larger $B\bar{B}$ backgrounds, which increase the experimental uncertainties, as shown in \Cref{tab:relative_uncertainty}. However, a trade-off exists with theoretical uncertainties, which reduce at lower energy thresholds as they become less dependent on the heavy-quark distribution function~\cite{Buchmuller:2005zv}; {\it e.g.,} Ref.~\cite{Misiak:2020vlo} reports an uncertainty of 5\% on the branching fraction computation for $E^B_\gamma>1.6~\gev$. Backgrounds arise predominantly from events with an energetic photon from $\pi^0\rightarrow\gamma\gamma$ decays. 
Remaining non-signal decays are subtracted using simulated background expectations.\par  The expected relative uncertainties on the branching fractions are shown in \Cref{tab:relative_uncertainty}. 
 The systematic uncertainty is driven by uncertainties on the residual background contamination, which is estimated to be 10\% (5\%) for the baseline (improved) scenario. The baseline scenario corresponds to current Belle II performance. The improved scenario relies on ongoing studies of  $\pi^0\rightarrow\gamma\gamma$ veto modeling. In the baseline scenario the precision becomes limited by the systematic uncertainties at approximately 10~\invab of data whereas in the improved scenario the statistical power of the Belle II sample will be exploited up to approximately 50~\invab. 
 Construction of relative quantities like asymmetries will offer further reduction of systematic uncertainties and enhanced reach. 
 {\bf Inclusive analyses of radiative $B$ decays will offer unique windows over non-SM physics} throughout the next decade.

\begin{table}[htpb!]
\begin{tabular}{cccccc}
\hline
\hline

Lower $E_{\gamma}^B$ threshold  & \multicolumn{4}{c}{Statistical uncertainty} & Baseline (improved) \\
                      & 1~\invab            & 5~\invab            & 10~\invab           & 50~\invab           &   syst. uncertainty                           \\     \hline
1.4~\gev                   & 10.7\%  &  6.4\% & 4.7\% & 2.2\% & 10.3\% (5.2\%) \\
1.6~\gev                   &  9.9\%  &  6.1\% & 4.5\% & 2.1\% &  8.5\% (4.2\%) \\
1.8~\gev                   &  9.3\%  &  5.7\% & 4.2\% & 2.0\% &  6.5\% (3.2\%) \\
2.0~\gev                   &  8.3\%  &  5.1\% & 3.8\% & 1.7\% &  3.7\% (1.8\%) \\
\hline\hline
\end{tabular}
\caption{\label{tab:relative_uncertainty} Projected fractional uncertainties of the $B\to X_s \gamma$ branching fraction measurement for various $E_{\gamma}^B$ thresholds. The systematic uncertainty is presented for a baseline scenario when the remaining background is known to the 10\% level, and an improved scenario, when the background is known to the 5\% level.
}
\end{table}

\subsubsection{\texorpdfstring{$B\to K^*\gamma$ and $B\to K\pi\pi\gamma$}{KstargammaKpipigamma}}
Exclusive radiative decays offer a complementary probe of non-SM physics in $b \to s$ gamma transitions. Experimentally they are more straightforward  than inclusive decays, but absolute rates suffer from larger theoretical uncertainties. The most important exclusive $b\to s\gamma$ decay is $B\to K^*(892)\gamma$ because of its relatively large branching fraction and low-multiplicity final state.  The most sensitive observables are relative quantities, such as ratios,  that suppress uncertainties associated with form factors~\cite{SM1,SM2,SM3, SM4}. They include the isospin asymmetry $$
	\Delta_{0+} = \frac{\Gamma(B^{0}\to K^{*0}\gamma) - \Gamma(B^{+}\to K^{*+}\gamma)}{\Gamma(B^{0}\to K^{*0}\gamma) + \Gamma(B^{+}\to K^{*+}\gamma)},$$ the \CP-violating asymmetries for $B^0$ and $B^+$ decay $A^{B^{+/0}}_{\CP}$, and their difference $\Delta A_{\CP} = A^{B^{+}}_{\CP} - A^{B^{0}}_{\CP}$. 
	Belle reported the current best results~\cite{Belle_paper}. They have $\mathcal{O(}10^{-2})$ precision, dominated by statistical uncertainties for the \CP violating asymmetries, and by the uncertainty on the ratio of $\Upsilon(4S)$ branching fractions between neutral and charged $B$ meson pairs ($f_{+-}/f_{00}$) for the isospin asymmetry.  
	

Expected statistical and systematic uncertainties of relevant observables are extrapolated from these results and presented as functions of luminosity in Table~\ref{table:observables_stat}. 
Belle~II will halve the systematic uncertainties with respect to Belle since (i) the leading source due to photon efficiency will be reduced from 2\% to below 1\%, as achieved by \babar\ and demonstrated in early Belle~II data; and (ii) the systematic uncertainties related to the fit bias for modes involving \piz will be halved thanks to an improved fit model.  The uncertainty on $f_{+-}/f_{00}$ will be reduced using unbiased measurements based on inclusive semileptonic decays~\cite{Jung:2015yma}.
While the precision on the \CP asymmetries will continue to improve until  surpassing the precision of the SM prediction~\cite{SM3}, $\Delta A_{\CP}$ will remain {\bf strongly sensitive to non-SM contributions up to the final data set size.}
Another powerful observable is the photon polarization, which is predominantly left-handed in the SM, but can acquire a right-handed component if non-SM contributions are present. The LHCb measurement of angular correlations in $B^{+}\to K^{+}\pi^{+}\pi^{-}\gamma$ decays showed strong evidence for polarization in a $b\to s\gamma$ process~\cite{LHCb_polarisation}. Belle~II will provide a 1\% measurement of the polarization parameter using this, and the complementary $B^{0}\to K^{+}\pi^{-}\pi^{0}\gamma$ decay~\cite{Bellee:2019qbt}. 




%

\begin{table}[!ht]
	\caption{Projected statistical and systematic (absolute) uncertainties of relevant observables from $B \to K^{*}\gamma$ decays. 
	}
	\label{table:observables_stat}
	\begin{center}
	\begin{tabular}{ c c c c c  c}
		\hline\hline
		
		Observable &  1 ab$^{-1}$ & 5 ab$^{-1}$ & 10 ab$^{-1}$ & 50 ab$^{-1}$ & Systematic uncertainty \\
		\hline
		$\Delta_{0+}(B\to K^{*} \gamma)$ &	1.3\% &	0.6\% &	0.4\% &	0.2\% & 1.2\% \\ 
		$A_{\it CP}(B^{0} \to K^{*0}\gamma)$ &	1.4\% &	0.6\% & 0.5\% &	0.2\% & 0.2\% \\
		$A_{\it CP}(B^{+} \to K^{*+}\gamma)$ &		1.9\% &	0.9\% &	0.6\% &	0.3\% & 0.2\% \\
		$\Delta A_{\it CP}(B\to K^{*}\gamma)$ & 2.4\% &	1.1\% &	0.7\% &	0.3\% & 0.3\%\\
		\hline\hline
		
	\end{tabular}
	\end{center}
\end{table}

\subsection{Leptonic {\it B} decays}

Purely leptonic decays $B\to \ell \nu_\ell$, $\ell=e,\mu,\tau$ are suppressed by the CKM matrix-element $|V_{ub}|$ and a helicity factor proportional to the squared lepton mass. The decay constant $f_{B}$ can be computed with a theoretical uncertainty of 0.7\% using lattice QCD~\cite{Aoki:2021kgd}, making these modes sensitive probes for SM extensions  involving an extended Higgs sector or leptoquarks, which would alter the decay rate. In addition to probing charged non-SM fields, these decays will enable complementary determinations of the CKM matrix element $|V_{ub}|$ that could be instrumental in elucidating the inclusive-vs-exclusive discrepancy. The $B\to \tau \nu_\tau$, $B\to \mu \nu_\mu$, and $B\to e \nu_e$ branching fractions are estimated at around $1 \times 10^{-4}$, $4 \times 10^{-7}$ and $9 \times10^{-12}$ in the SM, respectively, based on the global values $|V_{ub}| = (3.82 \pm 0.24) \times 10^{-3}$ and $f_{B} = 188 \pm 7$ MeV/$c^{2}$~\cite{Aoki:2021kgd}.
Analyses are challenging as these decays are rare and involve final states with missing energy from neutrinos and few tracks.  Global values are $\mathcal{B}(B\to \tau \nu_\tau) = (1.06 \pm 0.19) \times 10^{-4}$ and $\mathcal{B}(B\to \mu \nu_\mu) < 8.6 \times 10^{-7}$ and $\mathcal{B}(B\to e \nu_e) < 9.8 \times 10^{-7}$ at $90\%$ confidence level~\cite{Belle:2019iji,Belle:2006tbq}.  
{\bf Precision studies of these modes are exclusive to Belle~II} given their missing-energy signatures. Two-body decays $B\to \ell \nu_\ell$ with $\ell = e,\mu$ have a stable charged lepton in the final state, with monochromatic momentum in the rest frame of the signal $B$ meson, which provides for a distinctive feature to reconstruct the decay. The decay $B\to \tau \nu_\tau$ suffers from multiple neutrinos from both the $B$ and $\tau$ decays. The partner $B$ meson in the event is therefore reconstructed in specific decay modes using partner $B$ reconstruction. The  $B\to \tau \nu_\tau$ signal is extracted by inspecting the residual energy in the calorimeter, which peaks around zero for signal. 

\begin{figure}[htp]
\centering
\includegraphics[width=.65\linewidth]{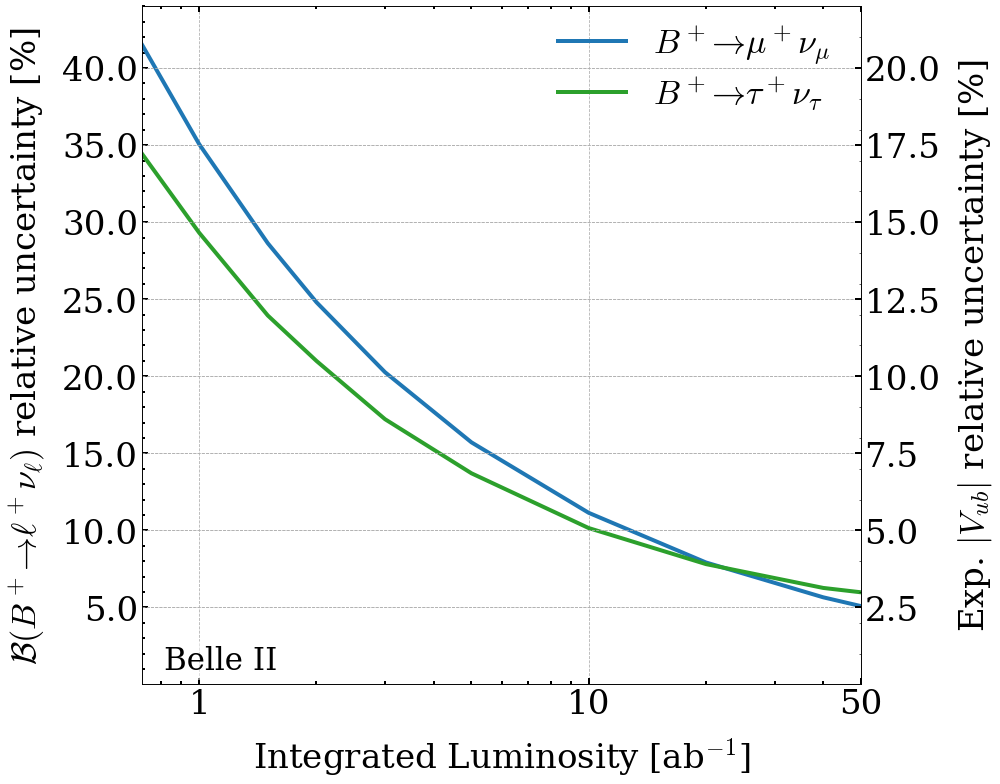}
\caption{Projection of uncertainties on the branching fractions ${\cal B} (B^+ \to \mu^+ \nu_\mu)$ and ${\cal B} (B^+ \to \tau^+ \nu_\tau)$. The corresponding uncertainty on the experimental value of $|V_{ub}|$ is shown on the right-hand vertical axis.}
\label{fig:btomunu_BV}
\end{figure}

  Baseline projections based on Belle results combined with demonstrated Belle II improvements (Figure~\ref{fig:btomunu_BV}) suggest that Belle~II will observe $B\to \mu \nu_\mu$ decays and improve significantly the precision on the $B\to \tau \nu_\tau$ rate.  Figure~\ref{fig:btomunu_BV} also shows the corresponding projections for the uncertainty on the branching fractions and $|V_{ub}|$ determinations for $B\to \mu \nu_\mu$ and $B\to \tau \nu_\tau$ decays (where the partner $B$ is reconstructed in its hadronic decay) as functions of integrated luminosity. Similar asymptotic precision is expected since the lower $B\to \mu \nu_\mu$ yield is compensated by a more straightforward reconstruction. At 50 ab${}^{-1}$ the $B\to \tau \nu_\tau$ precision will be limited by the $K_L$ veto efficiency, the calibration precision of the hadronic partner-$B$ reconstruction, and peaking backgrounds contributions. The $B\to \mu \nu_\mu$ will be limited by the continuum and $b \to u \ell \bar \nu_\ell$ background modeling precision. Introducing the benefits of inclusive partner $B$ reconstruction as demonstrated in the $B \to K^{(*)}\nu\nu$ analysis ~\cite{Belle-II:2021rof} will further the reach.   The ratio between the $B\to \tau \nu_\tau$ and $B\to \mu \nu_\mu$ and branching fractions, which is predicted in the SM with large precision, will offer a novel probe to test lepton-flavor universality.

\newpage

\section{Charm physics\label{sec:charm}}
Belle~II will have a broad charm physics program. A variety of measurements will provide numerous opportunities to test the SM and potentially uncover non-SM physics in ways that are complementary to the searches performed with $B$ decays. Decay-time-integrated measurements will allow for searches for direct \CP violation in several dozen decay modes, many of which (\eg, $D^0 \to \pi^0 \pi^0$, $D^0 \to \rho^0 \gamma$) can only be precisely measured at Belle~II. 
Decay-time-dependent amplitude analyses will be performed for multibody decays, many of which contain neutral particles such as $D^0 \to K^+ \pi^- \pi^0$ and $D^0\to \KS\pi^+\pi^-\pi^0$; this might provide additional sensitivity to charm mixing and mixing-induced \CP violation. The uniformity of the Belle~II acceptance and reconstruction efficiency as a function of final-state-particle kinematic properties and decay time makes such analyses straightforward. The efficient reconstruction of low-transverse momentum particles, and precise knowledge of the position and size of the interaction region, enable reconstruction of $D^*$-tagged rare charm decays having no decay vertex such as $D^0 \to \gamma \gamma$. The near hermeticity of the detector and known energy and momentum of the initial state enable reconstruction of charm decays with final-state neutrinos, such as decays involving $c \to u \nu \bar{\nu}$ transitions. Precise measurements of leptonic and semileptonic charm-strange decays such as $D_{s}^+ \to  \mu^+ \nu$ and $D_{s}^+ \to \phi \mu^+ \nu$, with input from lattice QCD~\cite{Boyle:2022uba}, will provide among the world's best direct knowledge of $|V_{cs}|$; alternatively, taking $|V_{cs}|$ from global fits to $B$-decay measurements and assuming CKM unitarity, they will provide measurements of decay constants and form factors that can be directly compared to lattice QCD calculations. A full program of charm baryon measurements will be pursued, including unique searches for \CP violation. Below we discuss in more detail a few measurements that highlight the strengths of Belle~II for charm physics.

\subsection{{\it CP} violation in \texorpdfstring{$D^{+,0}\to\pi^{+,0}\pi^0$}{Dpipi} decays}
The only observation of \CP violation in charm reported so far is the \CP-asymmetry difference between $\Dz\to\Kp\Km$ and $\Dz\to\pip\pim$ decays~\cite{LHCb:2019hro}. In contrast to $B$ decays, loop amplitudes in charm are severely suppressed by the Glashow-Iliopoulos-Maiani (GIM) mechanism~\cite{Glashow:1970gm}. Standard model \CP violation arises mostly from the interference of tree-level amplitudes, possibly associated to rescattering~\cite{Grossman:2019,Bediaga:2022sxw}. Rescattering amplitudes are challenging to compute and make the interpretation of the observed \CP violation ambiguous~\cite{Chala:2019}. Precise measurements of \CP asymmetries in other decay channels are crucial to understand the underlying dynamics. Cabibbo-suppressed decays such as $D^+ \to \pi^+\pi^0$ and $D^0 \to \pi^0\pi^0$ are particularly interesting due to their different isospin transitions compared to $D^0 \to \pi^+\pi^-$~\cite{Grossman:2012eb}. The $\pi^+\pi^0$ final state must have isospin $I=2$ to meet symmetry constraints, and thus the decay proceeds via a $\Delta I = 3/2$ amplitude. As the rescattering amplitude has $\Delta I = 1/2$, and the electroweak loop is highly suppressed, the decay proceeds essentially via a single tree amplitude. Thus, observing direct \CP violation in $\Dp\to\pip\piz$ would be a robust indication of non-SM physics. 
 A related probe for non-SM physics is provided by an isospin-based sum rule that relates branching fractions, \CP asymmetries, and total widths of $D^0\to\pi^+\pi^-$, $D^+\to\pi^+\pi^0$, and $D^0\to\pi^0\pi^0$ decays to discriminate whether observations of \CP violation in the individual channels are due to non-SM physics or not~\cite{Grossman:2012eb}. 
 The properties of the $D^0\to\pi^0\pi^0$ decay at the required level of precision will only be accessible at Belle II.

\begin{table}[!ht]
\begin{tabular}{lcccc} \toprule 
Int.~luminosity & 1\invab & 5\invab & 10\invab & 50\invab  \\
\midrule 
$\sigma_{A_{\CP}}(D^+\to\pi^+\pi^0)$ & 1.64\% & 0.74\% & 0.52\% & 0.23\%    \\
$\sigma_{A_{\CP}}(D^0\to\pi^0\pi^0)$  & 0.49\% & 0.22\% & 0.15\% & 0.07\%    \\
\bottomrule 
\end{tabular}
\caption{\label{tab:Acp4Charm} Expected statistical uncertainties on $A_{\CP}(D^{+,0}\to\pi^{+,0}\piz)$ as a function of Belle~II integrated luminosity. The projections are based on $D^{*+}$-tagged decays. 
}
\end{table}

 Table~\ref{tab:Acp4Charm} shows the expected sensitivity using $D^{*+}$-tagged decays reconstructed in the currently available Belle~II data, extrapolated to larger sample sizes. Only the statistical sensitivity is considered, as systematic uncertainties will be small.
 {\bf\boldmath Belle~II will dominate the precision on $A_{\CP}(\Dp\to\pip\piz)$ and will be the only existing 
experiment able to precisely measure \mbox{$A_{\CP}(\Dz\to\piz\piz)$}.}
 Additional sensitivity will come from untagged $D^+$ decays and from $\Dz\to\piz\piz$ decays where the $D^0$ flavor is inferred from the rest of the event or from flavor-specific $B \to D X$ decays~\cite{Belle-II:2018jsg}. 
Non-SM physics can generate \CP asymmetries in $D^{+,0}\to\pi^{+,0}\piz$ decays that are factors larger than observed in $\Dz\to\Kp\Km$ and $\Dz\to\pip\pim$ decays~\cite{Bause:2020}. With 50\invab, Belle~II will be the only experiment with the sensitivity to possibly observe \CP violation in $\Dp\to\pip\piz$.
Together with the unmatched precision expected on $A_{\CP}(D^0\to\pi^0\pi^0)$, and hence on the verification of the $\D\to\pi\pi$ sum rule~\cite{Grossman:2012eb}, {\bf\boldmath Belle~II will have unique potential to discover new sources of \CP violation in charm decays}.

\subsection{Rare charm decays} 
The rare decay $D^0\to\gamma\gamma$ is a $c\to u$ flavor-changing neutral-current process that will be uniquely probed at Belle~II. While its SM branching fraction is at most $10^{-8}$, it can receive 100-fold enhancements under various SM extensions~\cite{Greub:1996wn}. The most restrictive limit $\mathcal{B}(D^0\to\gamma\gamma)<8.5\times10^{-7}$ at 90\% CL~\cite{Belle:2015pzk} is just about two orders of magnitude above the SM prediction~\cite{Fajfer:2001ad, Burdman:2001tf}. Belle~II will probe a significant portion of the interesting non-SM parameter space, as reported in Table~\ref{tab:UL4BrDzTogg}. 

\begin{table}[!ht]
\begin{tabular}{ccccc}
\toprule 
Int. luminosity & 1\invab & 5\invab & 10\invab  & 50\invab \\ \midrule 
  $\mathcal{B}_{UL}^{90\%}(D^0\to\gamma\gamma)$ ($10^{-7}$)
            & 8.5 & 4.9 & 2.7  & 1.5   \\
\bottomrule 
\end{tabular}
\caption{\label{tab:UL4BrDzTogg}Expected upper limit on the branching fraction of $D^0\to\gamma\gamma$ decays as a function of Belle~II integrated luminosity.}
\end{table}

Another area where Belle II has distinctive access to non-SM physics are charm decays into final states involving neutrino pairs such as $D^{0,+}\to\pi^{0,+}\nu\bar{\nu}$, $D_s^+\to K^+\nu\bar{\nu}$, or $\Lc\to p\nu\bar{\nu}$. Since these decays are strongly GIM-suppressed, observing a significant rate could signal non-SM physics~\cite{Bause:2020xzj,Golz:2021imq}. 
Decays to final states with undetectable particles will be identified by reconstructing the partner charm hadron along with the particles arising from the fragmentation system, and then using energy-momentum conservation to determine the invariant mass of the remaining (recoiling) system~\cite{Belle:2016qek,Belle:2007wtt,Belle:2013isi}. This powerful technique enables {\bf searching for charm decays to final states with neutrinos, dark matter particles, axions, and other non-SM states, with sensitivities that will be unmatched by any other experiment in the next decade.}
 

\section{Quantum chromodynamics\label{sec:qcd}}

\subsection{Quarkonium, exotics, and hadron spectroscopy}
{\bf Belle~II offers unique possibilities for the discovery and interpretation of exotic multiquark combinations to probe the fundamentals of QCD.}

Difficulties in QCD calculations hinder accurate predictions of the spectra of hadrons. The interplay of experimental observations and semi-phenomenological effective models is needed. In recent years, experiments have observed several non-conventional bound states that could be explained as weakly-bound meson molecules, quark-gluon hybrids, tetraquarks, or pentaquarks. From initial observations, we are moving into an era where robust independent confirmation, precision knowledge, and discovery of partner states are essential to settle the questions raised by competing theoretical explanations.

Studies of quarkonium, the bound state of heavy quark-antiquark pairs ($c\overline{c}$, $b\overline{b}$), have been the gateway for the discovery of nearly all new multiquark states, via decays to and from conventional quarkonium particles. The $X(3872)$ \cite{PhysRevLett.91.262001} was the first of a growing alphabet of charmonium-like particles ({\it e.g.}, $X(3872)$, $Y_{c}(4260)$ \cite{PhysRevLett.95.142001}, $Z_{c}^{\pm}(3900)$ \cite{PhysRevLett.110.252001}, and several others) that also do not fit the well-established theoretical framework \cite{BRAMBILLA20201}. Analogous discoveries containing bottom quarks ({\it e.g.}, $\Upsilon(5S)$ decays to $Z_{b}^{\pm}(10610/50)$ \cite{PhysRevLett.108.122001}) indicate a similar unexplored family of particles in the bottomonium sector. The Belle~II experiment offers several unique opportunities in this domain. It will exploit 40 times more data than previous generation $B$-factories and, compared with hadron-collisions experiments, leverages a greater variety of quarkonium production mechanisms including $B$ meson decays, initial state radiation (ISR), double $c\overline{c}$ processes, two-photon processes, and direct production by changing collider center-of-mass energy \cite{Belle-II:2018jsg}. Belle~II is the only experiment with the ability to operate at a tuneable center-of-mass energy near the $\Upsilon(4S)$ resonance, providing direct access to multi-quark states containing bottom quarks. In addition, Belle~II's good efficiency for reconstructing neutral final-state particles opens the pathway for first observations of the predicted neutral partners of charged tetraquark states.

The Belle~II program on charmonium spectroscopy capitalizes on large, low-background samples and specific experimental capabilities. 
Efficient photon and $\pi^0$ reconstruction makes a lineshape measurement of $X(3872)$ via $D^0 \overline{D}^{0*}$ decays possible. This is more sensitive than current determinations in discriminating the quasi-bound or quasi-virtual nature of the $X(3872)$ as it sits right above the $D^0 \overline{D}^{0*}$ mass threshold where Belle~II has excellent experimental resolution~\cite{X3872toDDstarBelle}. A $3\sigma$ measurement of the width is expected with 10 ab$^{-1}$, becoming a solid observation with 25  ab$^{-1}$. In addition, the large production rate of $b\to c \bar c$ provides fertile ground to search for new exotic states, such as $c \bar c s \bar s$ tetraquark states, $X(3872)$ partners decaying to $D^{*0} \overline{D}^{*0}$, hybrid states in $1^{++} \otimes 0^{-+}$, and to characterize better existing signals via amplitude analysis. In addition, ISR production of charmonium and exotic states accesses a wide mass range complementary to the BESIII experiment. Unique to Belle~II is the ability to search for $Z$ states in both $B$-meson decays and direct production via ISR, which appear to produce two distinct sets of $Z_{c}$ states. Determination of the production rates and decay rates could provide crucial information to theorists to help understand the wave functions of the $Z$ states. Data sets of 10 (50)~ab$^{-1}$ integrated luminosity approximately correspond to an equivalent on-peak luminosity between $300-500$ $(1500-2500)$ pb$^{-1}$/10 MeV over a range of $3-5$ GeV, allowing detailed study of $Y_{c}$ states, decays to various $Z_{c}^{\pm}$ states, and reaching the threshold for $\Sigma_{c}\overline{\Sigma}_{c}$ production.
Large samples will also benefit studies of double-charmonium production of $J/\psi(1S)$ with an accompanying charmonium(-like) state \cite{PhysRevD.79.071101}, searches for states recoiling against charmonium of other quantum numbers (\emph{e.g.}, $\eta_{c}+c\overline{c}$ and $\chi_{cJ}+c\overline{c}$ are opportunities only available at Belle~II), and searches for states produced in two-photon processes such as $c\overline{c}s\overline{s}$ decaying to $J/\psi\phi$~\cite{PhysRevLett.104.112004}. Past experimental observations in these processes have generally been far from \emph{a priori} theoretical predictions, so expanding this sector with Belle~II is a unique physics opportunity.

Belle~II has the unique opportunity to explore bottomonium(-like) states by operating at center-of-mass energies around 10 GeV, where only small samples exist worldwide: $\mathcal{O}(10)$~fb$^{-1}$ at $\Upsilon(1S,2S,3S,6S)$, $\mathcal{O}(100)$ fb$^{-1}$ at $\Upsilon(5S)$, and typically less than  1~fb$^{-1}$ at intermediate points.
This opens a fruitful program, as demonstrated by previous discoveries at $e^{+}e^{-}$ colliders that yielded first observations of predicted bottomonia ($\eta_{b}(1S,2S)$, $h_{b}(1P,2P)$, and $\Upsilon(1D_{2})$) and unexpected four-quark states ($Z_{b}^{\pm}(10610,10650)$, $Y_{b}(10753)$) \cite{PhysRevLett.117.142001,Mizuk2019}. Collisions at energies below the $\Upsilon(4S)$ allow for testing non-SM predictions in $\Upsilon$ decays to invisible or lepton-flavor-violating final states \cite{PhysRevD.72.103508,SANCHISLOZANO201171,PhysRevD.94.074023}. Collisions at even higher energies, to cover the $\Lambda_{b}\overline{\Lambda_{b}}$ threshold and to potentially search for other new $Y_{b}$ states, are possible with future accelerator upgrades.

The first step in this program was taken in late 2021 with a special data collection in the region near 10.75 GeV.  Analysis has just started to explore the nature of the newly-discovered $Y_{b}(10753)$ state. The full impact of these non-$\Upsilon(4S)$ data yet to be realized. However, this special run demonstrated the capability of SuperKEKB and Belle~II to efficiently produce and collect high luminosity collisions at non-$\Upsilon(4S)$ energies. This offers great promise to conduct a comprehensive fine-step energy scan to disentangle the complicated nature of final-state cross sections and bottomonium(-like) states above the $B\overline{B}$ threshold, and to collect large statistics samples for similar studies at dedicated nearby energies.

The unexpected violation of OZI-suppression and heavy-quark spin-symmetry in the hadronic transitions of $\Upsilon(4S,5S,6S)$ to lower bottomonium states led to the discoveries of the $Z_{b}(10610)$ and $Z_{b}(10650)$ states, the only charged four-quark states containing $b$-quarks known to date. Revisiting $\Upsilon(6S)$ energies (${\approx}11$ GeV) with Belle~II will provide sufficiently large samples to probe the exotic content of the $\Upsilon(6S)$ state, with sufficient energy to access potential undiscovered $\Upsilon(2D,1F)$ bottomonium multiplets and other exotic states.

Extending the SuperKEKB center-of-mass energy beyond the current 11.2 GeV limit would access the production threshold for $\Lambda_{b}\overline{\Lambda_{b}}$ and $b\overline{b}g$ hybrids and unlock further possibilities for observation of higher energy bottomonium(-like) states and exotics. This will require upgrades to the accelerator (\emph{e.g.} additional RF accelerator components and stronger fixed steering magnets). While this effort is not foreseen in the current Belle~II program, it should be considered as an important future opportunity to probe the frontiers of hadronic spectroscopy.

\subsection{Constraining hadronic vacuum-polarization in muon {\it g-2} and opportunities for precision QCD in hadronization}
The contents of this section are explored in more detail in a dedicated Snowmass whitepaper~\cite{qcdWhitepaper}.

One of the most topical measurements in HEP is that of the gyromagnetic ratio g of the muon, which is usually parametrized as the “anomaly” $a_\mu = (g-2)/2$. The current experimental value (combining the BNL E821 result with the first result from the Fermilab $g-2$ experiment) differs from SM predictions based on dispersion relations by $4.2\sigma$, $a_\mu (\textrm{exp}) - a_\mu (\textrm{theory}) = (26.0 \pm 7.9) \times 10^{-10}$~\cite{PhysRevLett.126.141801,Aoyama:2020ynm}. This difference might be a sign of non-SM physics. The theory uncertainty is dominated by the leading-order hadronic contribution (HVP), which can be calculated from the cross section $\sigma(e^+e^- \rightarrow \textrm{hadrons})$ measured in $e^+e^-$ experiments. The result,  HVP=$(693.1 \pm 4.0) \times 10^{-10}$, is dominated by \babar\  and KLOE measurements of $\sigma(e^+e^- \rightarrow \pi^+\pi^-)$. However, the \babar\  and KLOE measurements notably differ. This difference introduces a systematic uncertainty of $2.8 \times 10^{-10}$~\cite{Davier:2019can}.  Belle II will perform these measurements with larger data sets, and at least comparable systematic uncertainty, to resolve this discrepancy.\par
Furthermore, Belle~II's multi-ab$^{-1}$ data set  will facilitate new approaches to suppress systematic uncertainties, particularly from particle identification. Although the specific systematic studies still need to be refined, preliminary estimates indicate that a reduction of the uncertainty on the HVP contribution to $g-2$ to 0.4\%, is within reach\cite{Belle-II:2018jsg}. This will match the expected experimental precision on $g-2$\cite{Belle-II:2018jsg,Aoyama:2020ynm}.\par 
In addition, Belle II will make precision measurements of the cross sections for other hadronic channels such as $\sigma(e^+e^- \rightarrow \pi^+\pi^-\pi^0)$, $\sigma(e^+e^- \rightarrow \pi^+\pi^-\pi^0\pi^0)$,  and $\sigma(e^+e^- \rightarrow K^+K^-)$. 
{\bf Belle~ II's operation at the highest luminosity $\boldsymbol{e^+e^-}$ collider, as well as its excellent particle-identification capabilities, places it in a unique position to further the studies of the HVP contribution to $\boldsymbol{g-2}$ in the next decade.}

The low-background environment of $e^+e^-$ annihilation exploited at unprecedented statistical precision will enable {\bf highly impactful tests} of transverse-momentum-dependent QCD evolution and factorization in jet and hadron production. In jet production, most non-perturbative dependencies vanish and quantities as the transverse momentum imbalance are known up to next-to-next-to-next-to-leading-log precision in the strong-coupling expansion~\cite{Gutierrez-Reyes:2019vbx}.  For $e^+e^-$ production, theory uncertainties are small, enabling precision tests. In hadron production, non-perturbative input is needed in describing, for instance, the interaction with the QCD vacuum. Here insights on evolution and factorization can be gained in conjunction with future electron-ion collider data.
In addition to their intrinsic value as probes of fundamental aspects of QCD, these studies will be {\bf essential to extract the three-dimensional nucleon structure from the upcoming experimental program at Brookhaven's Electron-Ion Collider}(EIC). Similarly, precise measurements of fragmentation functions from Belle~II will offer important inputs for the EIC and JLab  programs, as per the 2015 long-range plan for nuclear physics~\cite{Aprahamian:2015qub}.

Measurements of multidimensional correlations of momenta and polarizations of final-state hadrons during hadronization will further our understanding of soft QCD and will enable refinement and tuning of Monte-Carlo event generators at levels that may be instrumental to reach the precision needed to accomplish the LHC program. The lever arm in collision energy with respect to LEP data offers a robust basis for extrapolation to LHC energies. The Belle~II data set size will enable unique fully multidimensional measurements that capture the fuller picture of hadronization dynamics. Examples include the partial-wave decomposition of dihadron correlations or $\Lambda$ spin-momentum correlations corrected for feed-down~\cite{qcdWhitepaper}.

\section{Tau lepton physics\label{sec:tau}}

The $\tau$ is the sole lepton massive enough to decay both into leptons and quarks thus allowing probing non-SM physics in mass-dependent couplings associated with the third generation~\cite{Belle-II:2018jsg}. 
With $4.6 \times 10^{10}$ $\tau^-\tau^+$ pairs produced in 50~ab$^{-1}$ of electron-positron annihilation data~\cite{Banerjee:2007is} and decaying in kinematically constrained and low-background  conditions, Belle~II will pursue a program of measurements in $\tau$ production and decay of unprecedented precision and breadth. $\tau$ leptons are produced in pairs with known (up to radiative effects) center-of-mass energy and decay through electroweak mechanisms. They offer a rich environment for precision measurements of several fundamental SM parameters and are sensitive to broad classes of SM extensions. The Belle~II $\tau$-lepton program covers lepton universality tests, determination of fundamental SM parameters, and searches of non-SM interactions via lepton flavor violation, lepton number violation, and baryon number violation. {\bf Belle II will lead progress in $\tau$-lepton physics throughout the next  decade}.

\subsection{Lepton flavor universality}

Tau decays allow for high precision tests of the fundamental SM assumption that all three leptons have equal coupling strength ($g_\ell)$ to the charged gauge bosons of the electroweak interaction (charged-current lepton universality). A broad class of SM extensions violate lepton universality, such as two-Higgs-doublet model~\cite{Jung:2010ik} or singly charged-scalar singlet~\cite{Crivellin:2020klg} making searches for violation of lepton flavor universality attractive. While LHC measurements of lepton universality via $W^{\pm}$-boson decays are sensitive only to charged currents~\cite{ATLAS:2020xea, CMS:2022mhs} tests of lepton flavor universality via $\tau$ leptons offer additional sensitivity to non-SM contributions to weak neutral currents~\cite{Altmannshofer:2016brv, Bryman:2021teu}. \par Measurements typically consist in determining precisely branching-fractions ratios, such as $R_\mu \equiv \BRtautomunu$ to test $\mu-$e charged-current lepton universality $g_\mu/g_e$ or $\frac{\BFtautopinu}{\BFpimutwo}$ and $\frac{\BFtautoknu}{\BFKmutwo}$ for $\tau$--$\mu$ charged-current lepton universality $g_\tau/g_\mu$. 
Particle identification is the key experimental challenge since particle species distinguish between the relevant $\tau$ decay modes. Given the large sample sizes, precision is typically limited by the uncertainties in the corrections needed to match particle-identification efficiencies determined from simulation to those observed in data calibration-samples.    For instance, the most precise measurement of $R_\mu$ has 0.4\% precision, which propagates to 0.2\% precision on $g_\mu/g_e$,  dominated by the systematic uncertainty associated with lepton identification~\cite{BaBar:2009lyd}. Early Belle II data show that lepton-identification uncertainties comparable to those in Ref.~\cite{BaBar:2009lyd} are at reach.  In this baseline scenario, Belle~II would match the current best results, yielding a significant improvement in global precision. In an advanced scenario where an improved understanding of the detector and availability of more abundant and diverse calibration samples reduces lepton-identification uncertainties by a further factor of two, Bellle II would lead the $g_\mu/g_e$ determination.\par  
An important input to lepton-flavor universality tests are measurements of $\tau$ lifetime. The global value is dominated by the Belle result based on reconstructing both $\tau$ decays into three charged particles, $\tau(\tau)=(290.17 \pm 0.52 {\rm(stat)} \pm 0.33 {\rm(syst)}) \times 10^{-15}~\rm{s}$~\cite{Belle:2013teo}. The Belle~II data set size will significantly reduce the statistical uncertainties. The superior control of the vertex-detector alignment, demonstrated in recent charm lifetime measurements~\cite{Belle-II:2021cxx}, will reduce the dominant systematic uncertainty. The expected absolute precision of $0.2\times 10^{-15}~\rm{s}$ or better, will further improve the precision of $g_\tau/g_e$.
{\bf Belle~II will significantly improve the precision on inputs to lepton-flavor-universality-violating quantities yielding some of the most stringent constraints on non-SM deviations from charged current lepton universality.}  


\subsection{Cabibbo-angle anomaly}
Current global determinations of the CKM matrix parameter $|V_{us}|$ fall short of the prediction based on unitarity constraints by more than three standard deviations~\cite{HFLAV:2019otj}, generating the so-called Cabibbo-angle anomaly. The global precision of $|V_{us}|$ is dominated by measurements from kaon decays, which achieve 0.2\% (relative) precision combined with lattice QCD inputs.  Complementary information on $|V_{us}|$ is available from $\tau$-lepton decays, either using the ratio $R_{K/\pi} = \frac{\BR(\tau^-\to K^- \nu_\tau)}{\BR (\tau^-\to\pi^-\nu_\tau)}$ combined with lattice QCD inputs (exclusive determination), or the ratio of hadronic width into inclusive strange versus non-strange final states, corrected with SU(3)-breaking factors based on the operator-product expansion and finite-energy sum rules (inclusive determination). The current precision of $\tau$-based determinations is about two times worse than from kaon decays. Exclusive measurements are experimentally straightforward, provided that systematic uncertainties due to particle identification and background model are controlled at the desired precision. However, reliance on lattice-QCD inputs correlates the results with those from kaons. Inclusive measurements are challenging experimentally as they involve many correlated final states, but have the advantage of theory uncertainties independent of lattice QCD, thus providing genuinely independent consistency checks. However, they need exclusive measurements as input, and theory uncertainties are model-dependent and large. Reducing the latter would require  measurements of spectral density function of all strange final states. \par Belle II plans to pursue a dedicated suite of multiple measurements ultimately aimed at gaining a deeper insight on the sophisticated interplay between various experimental and theoretical approaches. Previous best results achieved 1.5\% precision on $R_{K/\pi}$~\cite{BaBar:2009lyd}, dominated by systematic uncertainties associated with particle identification and data-simulation discrepancies in background models.  Early Belle II data, with current 1\% uncertainties in hadron-identification-based corrections, demonstrate a competitive reach at 1.3\% precision. This may further improve to 0.9\% with halved hadron-identification uncertainties expected in future. Combined with planned measurement of the hadronic mass spectra and branching fractions of individual decay modes that will further reduce the theoretical uncertainties, this will improve the precision on $\tau$-based $|V_{us}|$ determinations providing {\bf unique and complementary input to gain insight on the Cabibbo-angle puzzle.}

\subsection{Charged lepton flavor violation} 

\begin{figure}[!ht]
    \centering
    \includegraphics[width=0.99\textwidth]{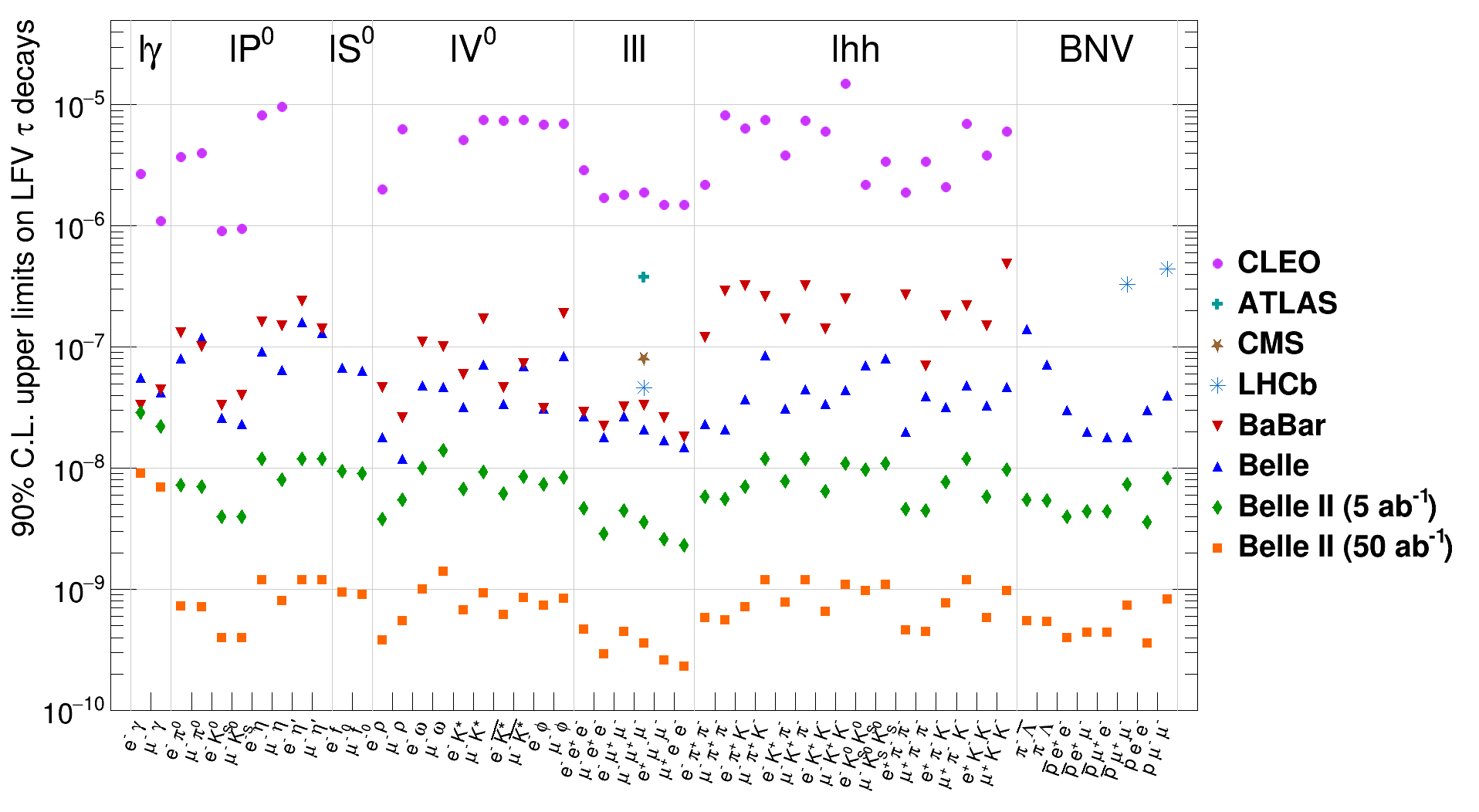}
    \caption{Current status of observed limits from CLEO, LHC and $B$-factory experiments~\cite{HFLAV:2019otj}, and Belle II projections of expected limits on charged-lepton flavor violating processes in $\tau$ decays.}
    \label{fig:TauLFV}
\end{figure}

No charged lepton flavor violation has been observed.  Minimal SM extensions that include right-handed neutrinos enable lepton-flavor violation, but through heavily suppressed mechanisms,  predicting branching ratios too small to be observed in current and foreseen experiments~\cite{Lee:1977tib,Petcov:1976ff}.  Extensions with non-SM interactions,  such as low-scale seesaw models~\cite{Cvetic:2002jy}, supersymmetric standard models~\cite{Ellis:1999uq, Ellis:2002fe, Dedes:2002rh, Brignole:2003iv, Masiero:2002jn, Hisano:2009ae,Fukuyama:2003hn},
little Higgs models~\cite{Choudhury:2006sq, Blanke:2007db}, 
leptoquark models~\cite{Davidson:1993qk},
$Z^\prime$ models~\cite{Yue:2002ja}, and extended Higgs models~\cite{Akeroyd:2009nu, Harnik:2012pb, Celis:2013xja, Omura:2015nja, Goudelis:2011un},
predict lepton-flavor-violation in $\tau$ decays at $10^{-10}$--$10^{-8}$ levels, which will be probed at Belle~II.
Several of these decay modes also include processes with lepton-number violation~\cite{LopezCastro:2012udb}  and baryon number violation~\cite{Hou:2005iu}.
The decays $\tau^-\to e^-\gamma$, $\tau^-\to\mu^- \gamma$ and $\tau^- \to \mu^- \mu^+ \mu^-$ are considered particularly promising as they are predicted at rates very close to the current experimental limits in a wide class of new physics models. Wrong sign decays, {\it e.g.}, $\tau^-\to\ell_j^-\ell_j^-\ell_i^+$ decays are also expected at rates only one order of magnitude below present bounds in some SM extensions~\cite{Pacheco:2021djh}. 

The current experimental status of charged-lepton-flavor-violating processes in 52 benchmark $\tau$-decay channels are shown in Figure~\ref{fig:TauLFV}~\cite{HFLAV:2019otj}, along with expected Belle II projections.
Projections are extrapolated for two illustrative scenarios of 5~\invab and 50~\invab based on expected limits in Belle analyses.  We consider the presence of irreducible backgrounds for $\tau^- \to \ell^-\gamma$ decays, thus assuming precision proportional to ${\cal{L}}^{-1/2}$. As a first approximation, we assume absence of relevant backgrounds for all other decays, approximating the precision as being proportional to ${\cal{L}}^{-1}$~\cite{Raidal:2008jk}. {\bf Belle~II will lead numerous  unique searches for charged-lepton-flavor violation} improving  current bounds by an order of magnitude or more, down to a few parts in $10^{-9}$--$10^{-10}$~\cite{Belle-II:2018jsg}.
Beam polarization~\cite{ChiralBelleWP} may further improve sensitivity since the dominant SM backgrounds depend on the polarization of electron beams whereas the uniform phase-space modeling of signal distributions does not~\cite{Hitlin:2008gf}. 

\subsection{Further non-SM probes}

\subsubsection{Anomalous electric and magnetic dipole moments} 
\label{tau-moments}

Precise measurement of the anomalous electric dipole moment (EDM, $d_\tau$) and magnetic moment ($(g-2)_\tau$) of the $\tau$ lepton are important tests of BSM physics possible at Belle~II.

The EDM of the $\tau$ lepton characterizes the time-reversal or \CP-violation properties at the $\gamma\tau\tau$ vertex. The SM predicts an extremely small value, $d_\tau \approx 10^{-37}$ $e$cm~\cite{Booth:1993af,Mahanta:1996er}, many orders of magnitude below any experimental sensitivity. This offers a powerful null test of the SM. Observation of a nonzero $d_\tau$ value would unambiguously indicate the presence of non-SM sources of \CP violation, which can enhance the EDM up to values of  $10^{-19}$ $e$cm ~\cite{Bernreuther:1996dr,Huang:1996jr}. 
Current best results are from a recent Belle measurement~\cite{Belle:2021ybo}, where the squared spin-density matrix of the $\tau$ production vertex is extended to include contributions proportional to the real and imaginary parts of the $\tau$ EDM.
Belle observed expectation values of the optimal observables~\cite{Diehl:1993br} consistent with zero with $10^{-18}~{e}\rm{cm}$ precision dominated by the uncertainty associated with discrepancies between data and simulations. Preliminary studies show that at Belle~II the data-simulation agreement improved significantly. Assuming conservatively a control on systematic uncertainties associated to momentum reconstruction, radiative effects, and charge asymmetries, similar to Belle,  Belle~II expects to probe the $\tau$ EDM at the $10^{-19}$ level or better with the 50~\invab data set. 
A proposed beam polarization upgrade at SuperKEKB will further improve the experimental sensitivity,  since the main contribution to the systematic uncertainty from the modeling of the forward-backward asymmetry in the detector response cancels for opposite beam polarization states~\cite{ChiralBelleWP}.
{\bf Belle~II will provide the most precise determination of the electric dipole moment of the $\tau$} with sensitivity at least an order of magnitude better than existing bounds~\cite{Ananthanarayan:1994af, Bernabeu:2006wf, Bernreuther:2021elu}.


Present deviations of the observed magnetic moment of the muon from its SM prediction make measurements of magnetic moment in $\tau$ leptons compelling, as the contribution to the anomalous magnetic moment of a lepton is enhanced by powers of the lepton mass. The experimental determination of the anomalous magnetic moment relies on the determination of the cross-section or partial widths for $\tau$-pair production, together with spin matrices or angular distributions of the $\tau$-decay products~\cite{Eidelman:2016aih, Bernabeu:2007rr, Chen:2018cxt}. The SM prediction for the magnetic moment of the $\tau$ lepton is around $1.1 \times 10^{-3}$~\cite{Eidelman:2007sb}, which becomes around $-2.7 \times 10^{-4}$ at Belle II energies~\cite{Crivellin:2021spu}.  The only experimental result is a bound with $10^{-2}$ precision~\cite{DELPHI:2003nah}.  
The experimental challenge is the control on the systematic uncertainties dominated by mismodeling in the simulation of backgrounds. The accompanying theoretical challenge is that an accurate calculation of higher-order QED corrections is needed. Assuming that these challenges will be met, preliminary studies based on the technique of optimal observables~\cite{Diehl:1993br} show that sensitivities sufficient to probe the SM prediction could be achieved. 
A possible future upgrade with beam polarization will provide access to the left-right asymmetry, which has enhanced sensitivity to the $(g-2)_\tau$~\cite{Eidelman:2016aih, Bernabeu:2007rr, Chen:2018cxt}. 
Measurement of asymmetries between left- and right-polarized beams benefit from cancellations of systematic uncertainties associated with  detector asymmetries, because the detector response does not change with beam polarization~\cite{Roney:2019til, ChiralBelleWP}.
{\bf Belle~II has the potential for a leading measurement of $(g-2)_\tau$} that will first probe the SM prediction. This would be of major interest in its own right, and even more if the muon anomaly inconsistency persists.

\subsubsection{Probing non-SM physics in G-parity-suppressed second-class currents}


Further probes of non-SM physics come from second-class-current processes. In the SM, branching fractions of $\tau \to \eta^{(\prime)} \pi \nu_\tau$ are suppressed by isospin-violating effects to a few parts in $10^{-5}$~\cite{leroy1978tau,weinberg1958charge,Nussinov:2008gx, Paver:2010mz, Descotes-Genon:2014tla, Escribano:2016ntp}. Large variations in estimates are due to choices in the modeling of the scalar form factor. This suppression offers an interesting case to search for non-SM physics, as the rates of $\tau^- \to \pi^- \eta^{(\prime)} \nu_\tau$ decays are greatly enhanced by the contribution of scalar currents from a putative extended Higgs sector~\cite{Jung:2010ik} or leptoquark bosons~\cite{Becirevic:2016yqi}. 
A precise measurement of the $\tau \to \eta^{(\prime)} \pi \nu_\tau$  branching fraction puts stringent constraints on non-SM scalar interactions, stronger than those from other low-energy observables~\cite{Garces:2017jpz}.

\begin{figure}[!ht]
    \centering
    \includegraphics[width=0.49\textwidth]{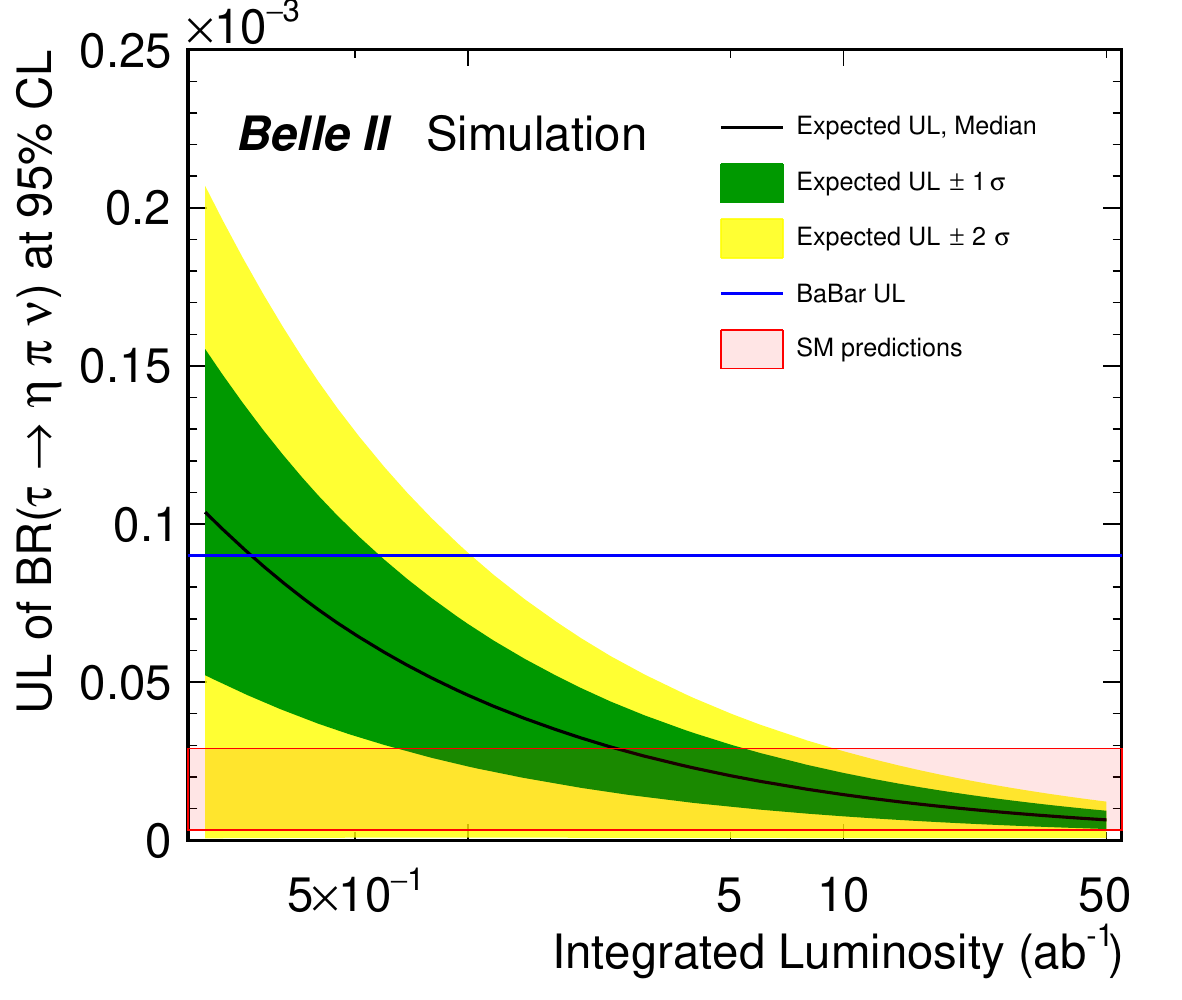}
    \includegraphics[width=0.49\textwidth]{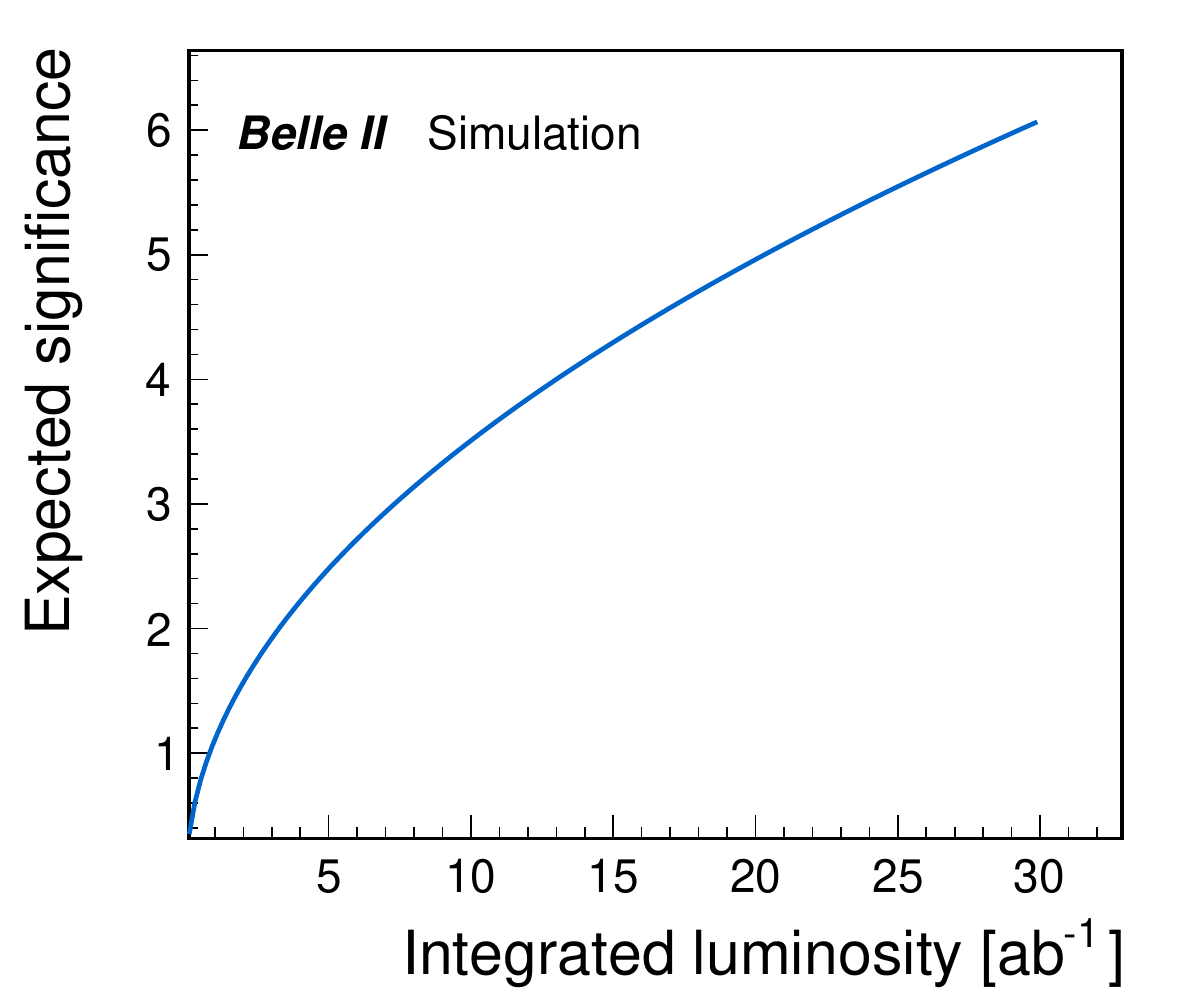}
    \caption{(Left panel) Expected upper exclusion limits on $\mathcal{B}(\tau^- \to \eta\pi^- \nu_\tau)$ at the 95\%~C.L., as  functions of data set size assuming absence of signal. The red band covers the most recent SM estimations for the branching ratio~\cite{Nussinov:2008gx, Paver:2010mz, Descotes-Genon:2014tla, Escribano:2016ntp}. (Right panel) Expected signal significance assuming a branching ratio of $1.67 \times 10^{-5}$ from~\cite{Escribano:2016ntp}.}
    \label{fig:scc_sensitivity_Belle2}
\end{figure}

Decays $\tau^- \to \eta^{(')} \pi^-\nu$ have not been observed because of significant backgrounds from other $\tau$ lepton decays~\cite{delAmoSanchez:2010pc, Hayasaka:2009zz}. The Belle~II sample size will allow a systematic study of other SM decays $\tau^- \to \eta + X$, providing the information needed to discriminate between isospin-violating $\tau^- \to \eta^{(')} \pi^-\nu$ decays and the isospin-conserving modes. Figure~\ref{fig:scc_sensitivity_Belle2} shows the expected upper limits for the branching ratio assuming that no significant signal is observed, and the expected significance assuming a SM signal~\cite{Escribano:2016ntp}. A precision measurement, accompanied by improved theoretical knowledge of the scalar form factor, 
will set {\bf stringent bounds on charged Higgs exchange competitive to those obtained from $B^-\to\tau^-\nu_\tau$ data}, even if no excess is seen over second class current predictions~\cite{Garces:2017jpz}. Following the observation of the $\tau^- \to \eta \pi^- \nu_\tau$ decay at Belle~II, the measurement of the hadronic spectrum and angular distributions will help in  disentangling non-SM physics from SM isospin-breaking effects.

\section{Direct searches for light non-SM physics and Dark Sector studies\label{sec:dark}}

The existence of dark matter~(DM) has been indirectly established in astrophysics. Identifying particles that explain the properties of DM is a chief goal of particle physics.  Strong exclusion constraints on weakly coupled DM at the electroweak scale motivate ``light" DM scenarios, with MeV--GeV masses for DM particles and mediators.
In dark sector models, DM and mediators are gauge singlets, and only mediators, which could be scalar, pseudo scalar, vector and fermion, couple to both SM and dark sector particles. Only a few interactions are compatible with SM symmetries, and are those mediated by dark photons, Higgs, neutrinos, and axions. Belle~II has unique or world-leading reach in searches for mediators at the MeV--GeV scale owing to high intensity collisions at 10~GeV center-of-mass energy.

\subsection{Axions and axion-like particles}
The QCD axion emerges from the Peccei-Quinn mechanism for solving the strong \CP problem. The mass of the axion could be large, around the MeV--GeV scale, if additional sources contribute to the mass. This is an attractive scenario for avoiding the `axion quality problem'~\cite{Kamionkowski:1992mf,Holman:1992us,Barr:1992qq,Ghigna:1992iv}. Since the axion naturally couples to gluons via $aG\tilde{G}$, a heavy axion could be produced in $B \to K a$ decays and decay hadronically. The decay of such an axion is expected to be displaced (prompt) for axion mass below (above) the $\eta\pi\pi$ threshold. Belle~II has unique sensitivities to both signatures~\cite{Chakraborty:2021wda,Bertholet:2021hjl}.

\begin{figure}[!ht]

    \centering
    \includegraphics[width=.45\linewidth]{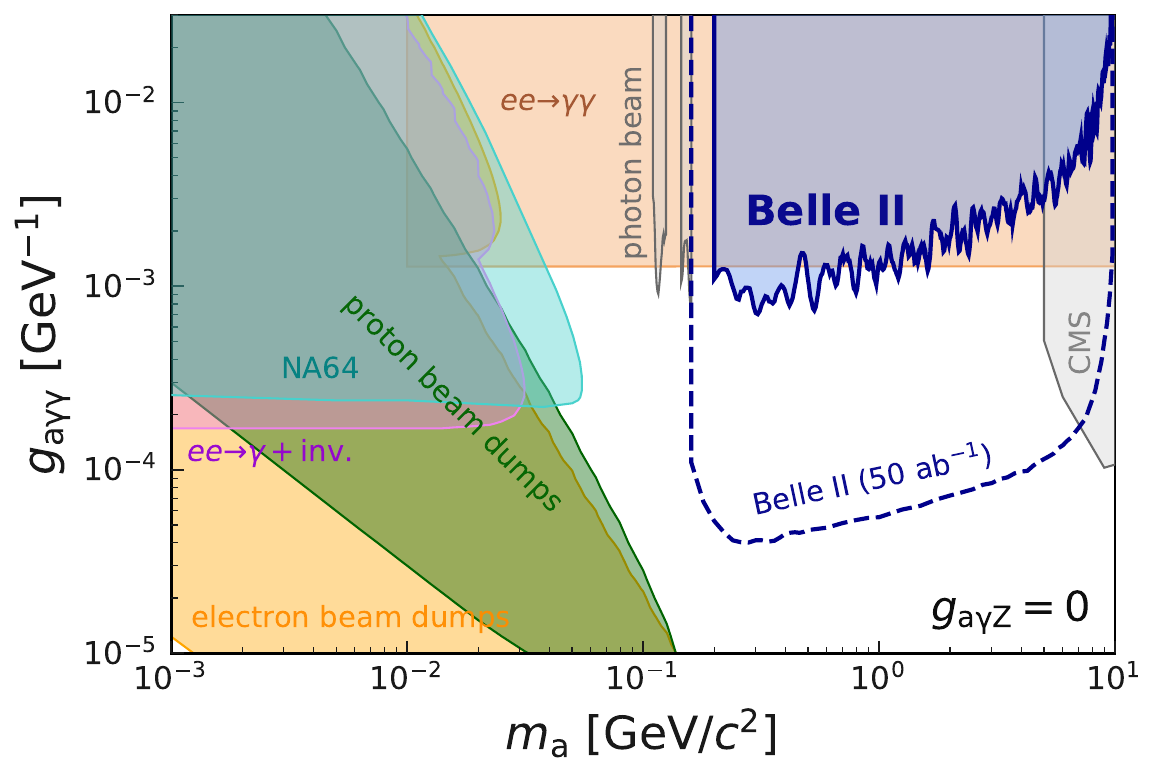}
    \caption{Expected Belle II sensitivity for ALPs with a three photon signature (blue, dashed) taken from \cite{Dolan:2017osp}. Sensitivity is expressed in terms of the coupling constant as a function of axion-like mass. Previous constraints from Belle II, electron beam-dump experiments, $e^+e^- \rightarrow \gamma +$ invisible, proton beam-dump experiments, $e^+e^- \rightarrow \gamma \gamma$, a photon-beam experiment, and heavy-ion collisions are taken from \cite{Belle-II:2020jti} and references therein.}
    \label{fig:ALPs}
\end{figure}

In analogy with the QCD axion, axion-like particles~(ALPs or $a$) are postulated pseudoscalar particles originating from the global-symmetry breaking of a general SM extension.  Contrary to the QCD axion, relations of ALPs masses and decay constants are unconstrained thus allowing a much broader parameter space.
Since ALP couplings to SM gauge bosons are generic, for simplicity we assume that ALPs only couple to diphotons. Hence they would be produced at Belle~II via $e^+e^- \to a \gamma$
and decay promptly to $a \to \gamma \gamma$ or be long-lived, depending on the mass and coupling $g_{a\gamma\gamma}$~\footnote{Two photon production $e^+e^- \to e^+e^-a$ and $B$ decays $B \to K a$ are also possible. The latter relies on ALPS coupling to photon and $W$ bosons.}.
The signatures are either three photons with total energy consistent with the center-of-mass energy or a single photon with large missing energy, if the ALP is too long-lived to decay inside the detector. The long-lived signature is the same as that of the invisible dark photon mentioned in Section~\ref{sec:DP}.
Belle~II will search for both signatures by reconstructing the axion from the diphoton mass or the recoil mass.
Results of the first Belle~II ALPs search~\cite{Belle-II:2020jti} offer a solid baseline to extrapolate future projections.
Figure~\ref{fig:ALPs} shows the first Belle II result ~\cite{Belle-II:2020jti} and the expected sensitivities for 50~ab${}^{-1}$~\cite{Dolan:2017osp} in the three-photon case only.
Expectations for single photon signatures are not shown here because the involved mass region is affected by a cosmic ray background whose characterization is yet to be completed (see also section~\ref{sec:DP}). 
{\bf Belle~II has unique sensitivity with three-photon signature in parameter-space regions unaccessible to beam-dump experiments.} 

\subsection{Dark photon}
\subsubsection{Minimal dark photon model}
\label{sec:DP}
\begin{figure}[!ht]
    \centering
    \includegraphics[width=.495\linewidth]{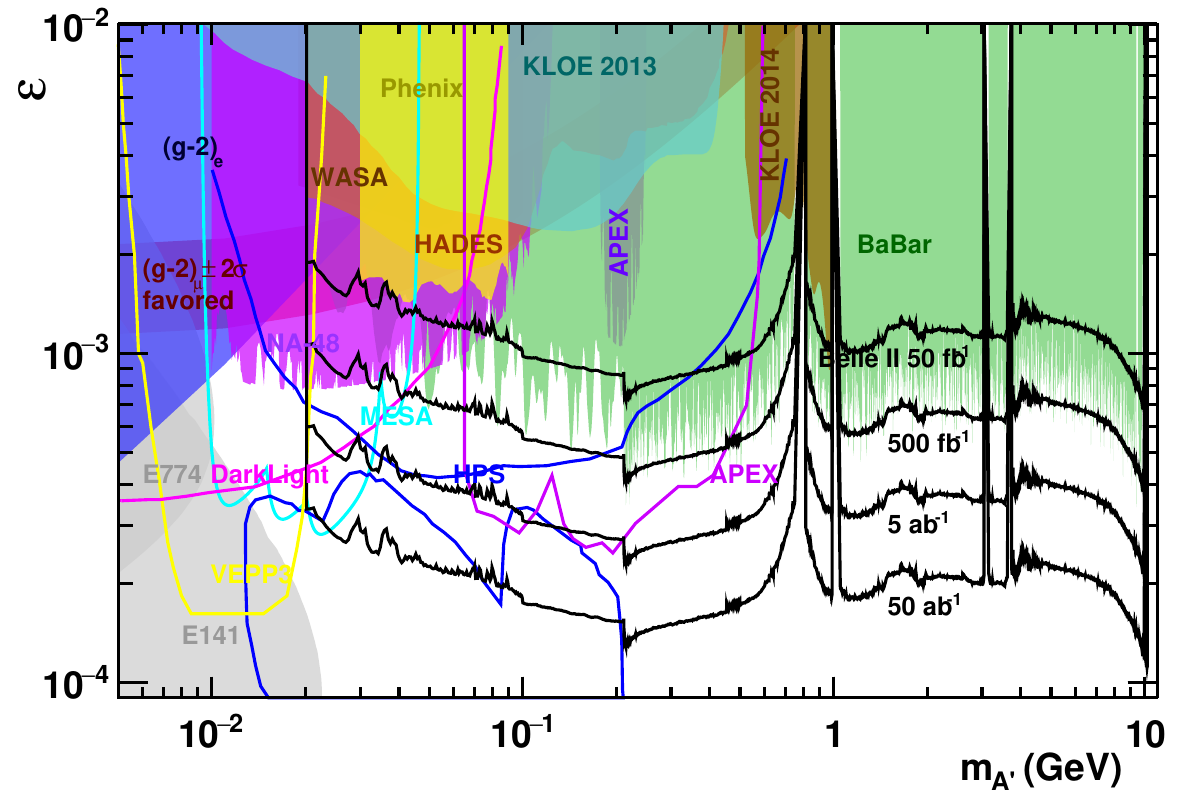}
    \includegraphics[width=.495\linewidth]{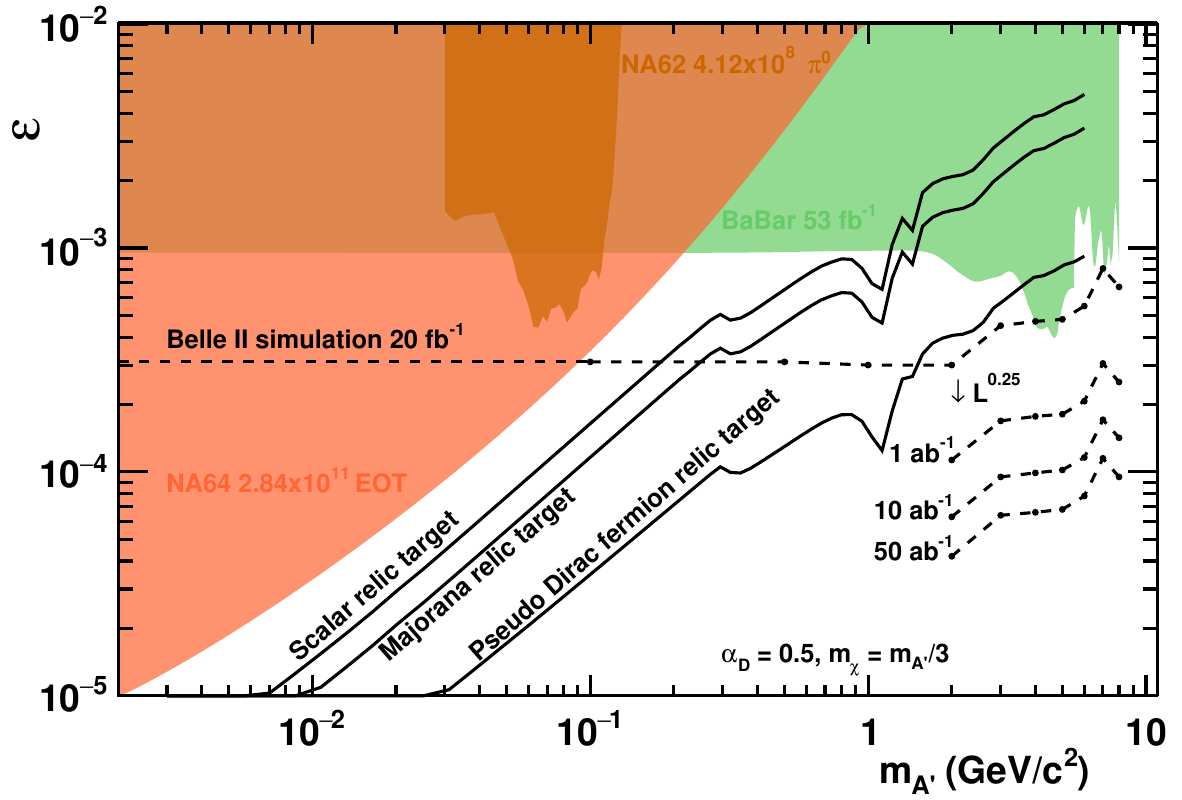}
    \caption{Projected sensitivities in terms of 
    the kinetic-mixing parameter $\epsilon$ as function of dark-photon mass in (left planel) visible  and (right panel) invisible signatures.}
    \label{fig:DP}
\end{figure}

A dark photon $A'$ is the postulated vector gauge boson from extra dark $U(1)$ symmetry and a vector portal mediator that couples to SM particles via kinetic mixing to SM photon~($\varepsilon$).  The dark photon can couple to the dark particles with dark $U(1)$ charge.
When a pair of the lightest dark sector particles~$\chi$ are lighter than dark photon, the dark sector particle $\chi$ could be a viable DM candidate. The dark photon can be produced via kinetic mixing $e^+e^- \to \gamma \gamma^* \to \gamma A'$. Belle~II has unique reach on the dark photon through its visible and invisible signatures. The visible signatures are decays to a charged-particle pair (electron, muon, pion, or kaon) via kinetic mixing; the invisible signature is a single photon with large missing energy. Figure~\ref{fig:DP} shows the expected sensitivities for visible and invisible dark photons.
We extrapolate visible dark-photon sensitivities from the existing BABAR limits, taking into account the larger drift chamber radius, which improves the mass resolution, and doubled trigger efficiency in the electron final state. The improved trigger is expected to become available in the barrel region of the calorimeter. We extrapolate  invisible dark-photon sensitivities using Ref.~\cite{Belle-II:2019usr} as a baseline and assuming statistical scaling with the luminosity. Projected sensitivities are not shown below 2 GeV, since that region is affected by a cosmic ray background whose characterization is yet to be completed. We also assume that a single photon trigger with comparable performance as the 2021 trigger will operate through the full data taking despite higher backgrounds.
The photon energy threshold currently applied 0.5~GeV corresponding to recoil mass of 10.1~GeV on $\Upsilon(4S)$ will be increased to 2.0~GeV~(8.3~GeV recoil mass) for higher luminosity to avoid exceeding the level~1 trigger limit of 30~kHz.
For visible dark photons, Belle~II will improve sensitivity on the kinetic mixing $\varepsilon$ by a factor of three over current best results. One order of magnitude better sensitivity at GeV masses will be achieved on the invisible signature, thus constraining regions of the parameter space for dark-sector models consistent with the observed relics density. Sensitivity at lower masses will depend crucially on capabilities to impose stringent vetoes on cosmic rays, to use the muon and $K^0_L$ detectorfor vetoing, and to maintain trigger efficiency high under future beam-background conditions.

\subsubsection{Extended dark photon models}
An hypothetical $A'$ may acquire mass through a spontaneous symmetry breaking of dark Higgs potential that introduces a physical dark Higgs boson $h'$. The mass of the dark Higgs can be either larger or smaller than the dark photon mass. Low-mass dark Higgs can be produced at SuperKEKB via the dark Higgsstrahlung process $e^+e^- \to \gamma^* \to A'^* \to A'h'$~\cite{Batell:2009yf}, with cross section proportional to $\varepsilon^2\alpha_D$, where $\alpha_D$ is the coupling constant of the dark gauge sector. 
If the dark Higgs is less massive than the dark photon and any other dark-sector particle, the dark Higgs is stable and invisible in the range of the detector. 
Belle~II searches for dark Higgs $h'$ with the signature $e^+e^- \to \mu^+ \mu^- + {\rm nothing}$ if the $A'$ decays to dimuon. 
A counting technique is used, relying on current Belle~II results~\cite{Rad_moriond2022}  and simulated background.  Figure~\ref{fig:lmultau} shows the expected sensitivities for the dark Higgsstrahlung search. Systematic uncertainties are assumed  to be at the same present level of 2\% both for signal and background.{\bf The Belle~II sensitivity for this topology is unique and limited by sample size even at the highest projected integrated luminosities}. Dark Higgs particles more massive than dark photons will be also searched for in six-fermion final states.

\begin{figure}[!ht]
    \centering
    \includegraphics[width=.495\linewidth]{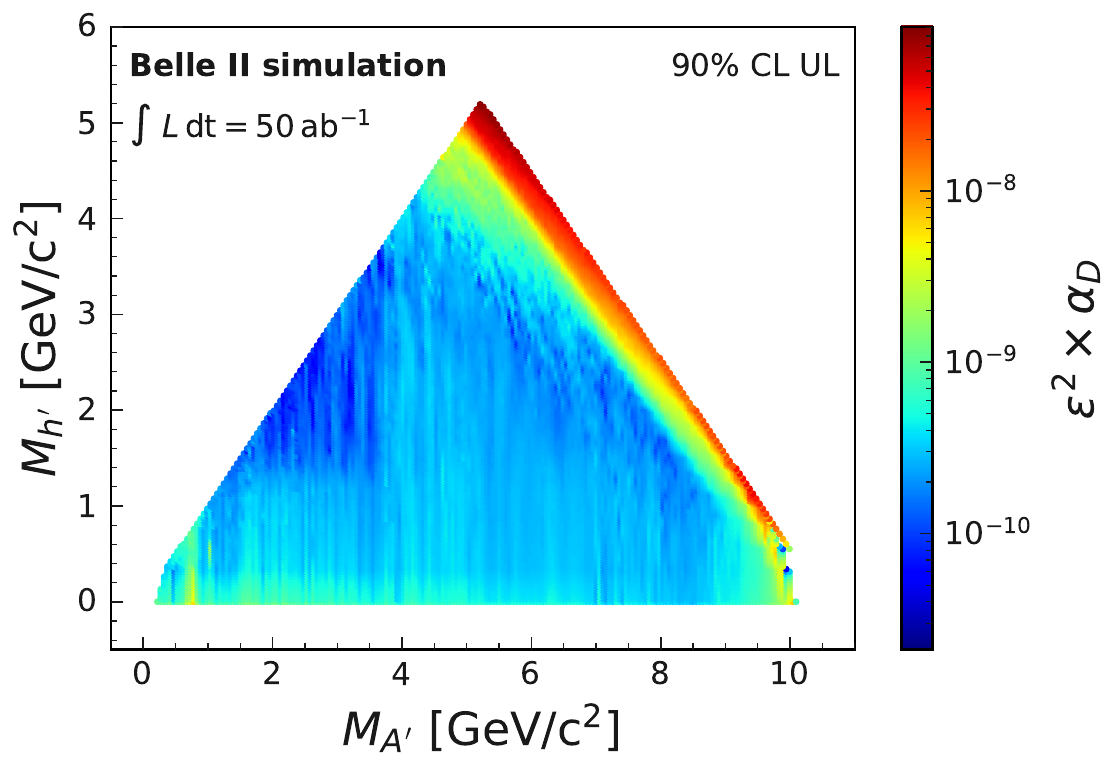}
    \includegraphics[width=.495\linewidth]{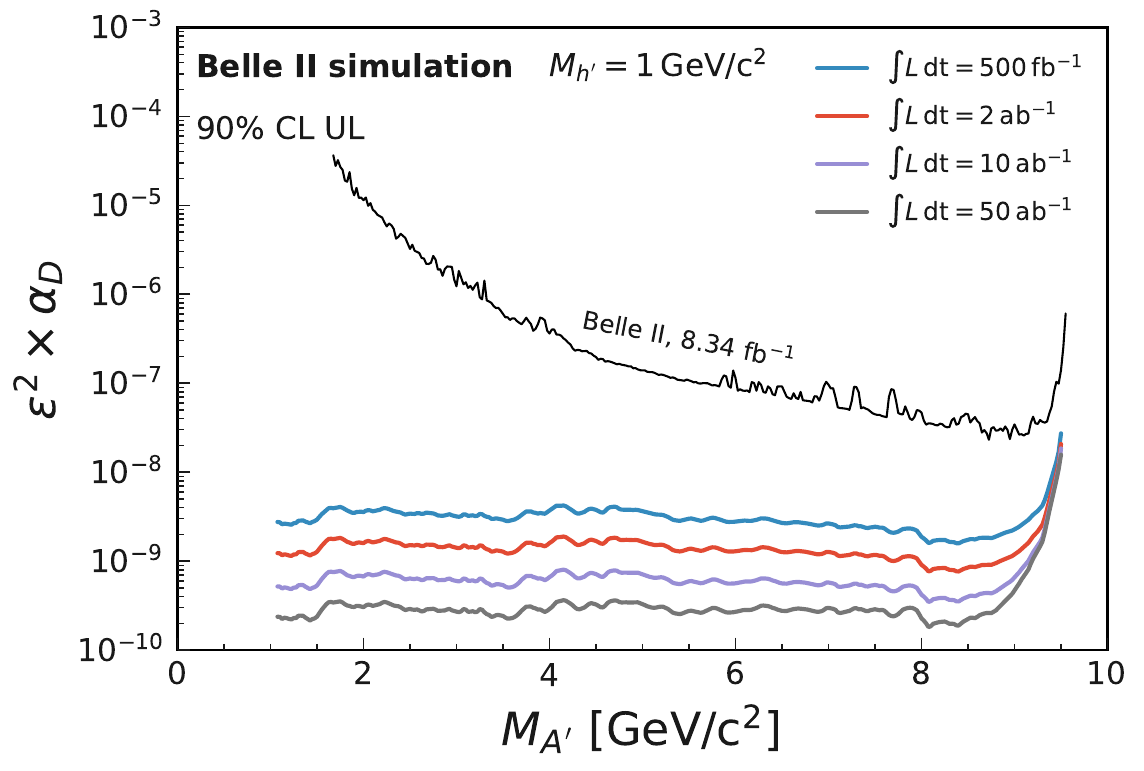}
    \caption{Sensitivity for dark Higgs boson~$h'$. (Left panel) upper limits on cross sections as functions of $m_{A'}$ and $m_{h'}$ with 50~ab${}^{-1}$ data. (Right panel) upper limit on 
    $\varepsilon^2\alpha_D$ for $m_{h'}=1~{\rm GeV}$ as a function of $m_{A'}$.
    }
    \label{fig:DH}
\end{figure}

\subsection{{\it Z'} in an \texorpdfstring{$L_{\mu}-L_{\tau}$}{LmuminusLtau} model}
\begin{figure}[!ht]
    \centering
    \includegraphics[width=.595\linewidth]{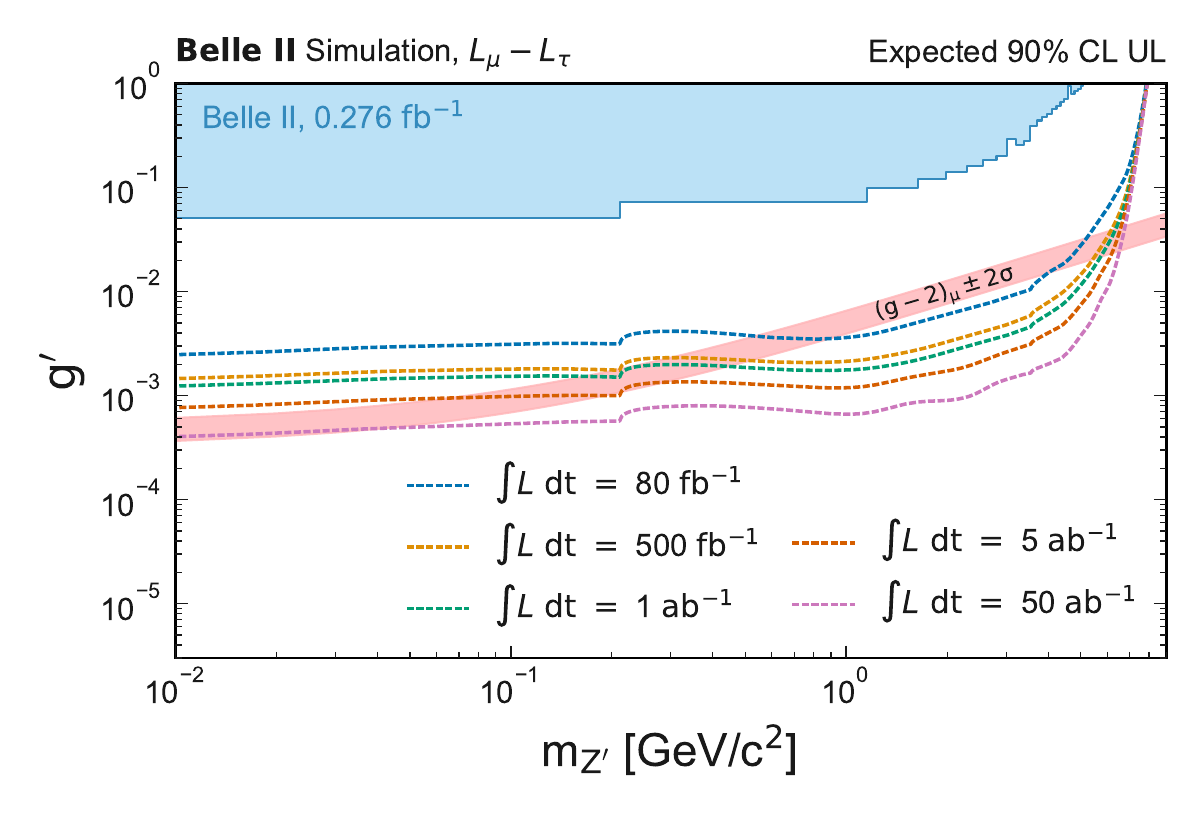}
    \caption{Sensitivity in terms of coupling constant as function of mass for a fully invisibly-decaying $Z'$ in a $L_{\mu}-L_{\tau}$ model. The blue band represents the excluded region by Belle~II~\cite{Belle-II:2019qfb}; the red band represents the parameter space relevant for explaining the observed muon $g-2$ anomaly.}
    \label{fig:lmultau}
\end{figure}

A particularly economic SM extension that could explain the muon $g-2$ anomaly~\cite{Muong-2:2006rrc, Muong-2:2021ojo} implies an anomaly-free addition of a new $U(1)_{L_{\mu}-L_{\tau}}$ gauge symmetry~\cite{He:1991qd}, where $L_{\mu}$ and $L_{\tau}$ are the lepton family numbers.  The new vector gauge boson $Z'$ couples to second and third generation leptons, $\mu, \nu_{\mu}, \tau$ and $\nu_{\tau}$, with the new coupling constant of $g'$. At Belle~II, an hypothetical $Z'$ can be radiated off muons, $e^+e^- \to \mu^+ \mu^- Z'$. If the couplings of $Z'$ to dark-sector particles are strong, invisible decays can dominate.
Belle~II searches for the $Z'$ with the same signature $e^+e^- \to \mu^+ \mu^- + {\rm nothing}$ as used in the dark Higgs search. Figure~\ref{fig:lmultau} shows expected sensitivities for $Z'$ in $L_{\mu}-L_{\tau}$ model.  {\bf The parameter space relevant for the muon $g-2$ anomaly will be nearly completely probed with 50~ab$^{-1}$}.$Z' \to \mu^+ \mu^-$ decays are also important for $m_{Z'} > 2m_{\mu}$ and will be searched for.

\subsection{Long-lived signatures}
\subsubsection{Inelastic Dark Matter}
\begin{figure}[!ht]
    \centering
    \includegraphics[width=.59\linewidth]{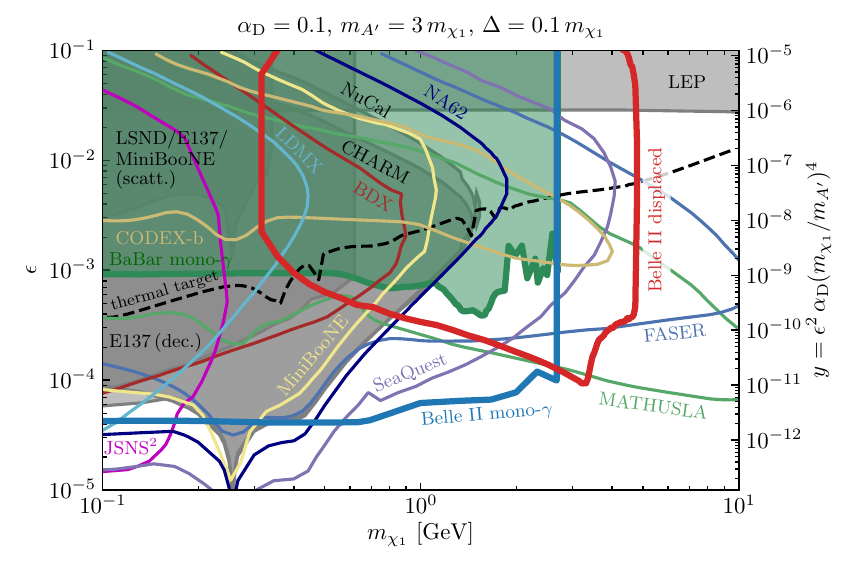}
    \caption{Projected sensitivity to iDM particles in terms of kinetic-mixing parameter 
    $\epsilon$ as a function of the scalar fermion mass. Projections refer to 50~ab${}^{-1}$ and (solid, red curve) displaced  or (solid, blue curve) monophoton signatures. The dark, dashed curve represents the observed relics abundance.}
    \label{fig:iDM}
\end{figure}

In a possible dark sector with numerous dark particles, if the mass difference between lightest and next-to-lightest dark particles~($\chi_1$ and $\chi_2$) is small and an inelastic transition between them is mediated by a dark photon $A^{\prime}$~\cite{Hall:1997ah,Tucker-Smith:2001myb}, $\chi_2$ becomes long-lived and $\chi_1$ is called inelastic dark matter~(iDM). This model satisfies the relics density with thermal freeze-out and is compatible with constraints from direct detection experiments. At Belle~II, the dark particles can be produced via $e^+e^- \to \gamma A' \to \gamma \chi_1 \chi_2$ if the $A'$ mass is higher than the summed $\chi_1$ and $\chi_2$ masses. The long lived $\chi_2$ decaying to $e^+e^-\chi_1$ is a distinctive  signature where the $e^+e^-$ system forms a displaced vertex and does not-point to the interaction point. 
Even with 50~ab${}^{-1}$, this search is background free thus the sensitivity is scaled with luminosity. 
The single photon signature is also usable when the $\chi_2$ escapes before decaying or the decay products are too soft to be observed.
Figure.~\ref{fig:iDM} shows the sensitivity for iDM with 50~ab${}^{-1}$~\cite{Duerr:2019dmv}. 
{\bf Belle~II will cover a major fraction of the region compatible with thermal relics.}

\subsubsection{Dark scalar}
\begin{figure}[!ht]
    \centering
    \includegraphics[width=.45\linewidth]{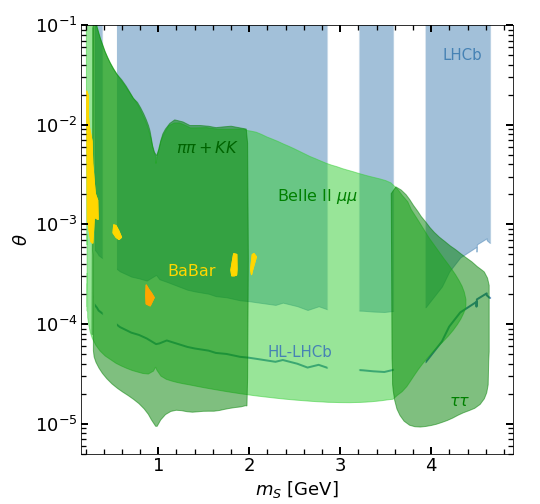}
    \caption{Projected $B \to K S$ sensitivity in terms of 
    $\theta$ as a function of visible $S$ mass for 50~ab${}^{-1}$.}
    \label{fig:darkscalar}
\end{figure}

In the minimal extension of a scalar mediator model, a real singlet scalar field $\phi$ and a dark fermion $\chi$ should be introduced. 
A new scalar field $\phi$ can mix with the neutral component of Higgs field with a mixing angle of $\theta$ and then physical dark scalar $S$ and observed Higgs boson $h$ appear after the symmmetry breaking~\cite{Krnjaic:2015mbs,Matsumoto:2018acr}. 
The dark scalar $S$ can couple to the SM fermion $f$~(dark fermion $\chi$) with strength proportional to $y_{\chi} \sin \theta$ ($y_f \cos \theta$), where $y_f$~($y_{\chi}$) is the Yukawa coupling of the $f$~($\chi$).
Since the couplings to heavier SM fermions should be larger, the dark scalar can be produced from $B$ meson penguin decays coupling to internal top quarks, $B^+ \to K^+ S$, for $m_S < m_B^+ - m_K^+$. 
The dominant decays are invisible $S \to \chi \bar{\chi}$ for $m_S > 2 m_{\chi}$. Visible decays of $S \to f\bar{f}$ are dominant for $m_S < 2 m_{\chi}$ and the dark scalar could be long-lived with small mixing angle\footnote{This signature is also sensitive to the ALPs which couple to fermions or W bosons, $B \to K a$ followed by $a \to f\bar{f}$.}.  Belle~II will search for the dark scalar in both decays.
Figure~\ref{fig:darkscalar} shows the expected sensitivity with 50~ab${}^{-1}$~\cite{Filimonova:2019tuy} for visible decays of $S \to \mu^+\mu^-,$ $\pi^+ \pi^-$, $K^+ K^-$ and $\tau^+ \tau^-$. Belle~II can reach mixing angle $\theta \approx 10^{-5}$ and access the region consistent with the thermal relics~\cite{Matsumoto:2018acr}.

\section{Precision electroweak physics with SuperKEKB polarization upgrade\label{sec:ewk}} 

Polarized $e^-$ beams open up new windows of exploration through measurement of the left-right asymmetry $(A_{\it LR})$ of the fundamental electroweak SM  parameter $\sin^2\theta_W$. A beam polarization upgrade has been proposed to pursue this unique precision program, along with the rest of the core physics program. This is described in more detail in an accompanying Snowmass 2021 Whitepaper~\cite{ChiralBelleWP}. $\tau$-lepton polarization measurements would provide precise ($< 0.5\%)$ determinations of beam polarization, required for  left-right asymmetry measurements, 
With 20~ab$^{-1}$ of beam polarized data, the precision on $\sin^2\theta_W$ measured with  $\tau$s, as well as with muon, electrons, b-quarks and c-quarks, will be comparable or better than the current world average.  As  $\sin^2\theta_W$ is  measured with multiple fermions, these studies will produce neutral-current lepton universality measurements of unprecedented precision. 
{\bf No other experiment, currently running or planned, can perform such precision tests of vector coupling universality in neutral currents~\cite{ChiralBelleWP}}.

Moreover, SuperKEKB  will yield the unique possibility of probing “dark forces” that can serve as portals between baryonic and dark matter. SuperKEKB with polarization complements other measurements as it is uniquely sensitive to a parity violating light neutral gauge boson in the dark sector ($Z'$) under various mass and coupling scenarios, including models where $Z'$ couples more to the 3rd generation via mass-dependent couplings. For example, a 15 GeV $Z'$ would cause a shift in the measurement of $\sin^2\theta_W$ in the energy region where SuperKEKB with polarized beams may have the best potential for discovery~\cite{Davoudiasl:2015bua}. With polarized beams SuperKEKB can also probe parity violating couplings of new heavy particles that couple only to leptons, complementing electroweak studies at the LHC. Further details are discussed in the Ref.~\cite{ChiralBelleWP}.

In addition to the precision measurements of the weak mixing angle at 10~GeV, this `Chiral Belle' physics program with polarized beam electrons  also enables the measurement of $(g-2)_\tau$ at an unprecedented and unrivaled level of precision as discussed in Section~\ref{tau-moments}.  
 Other  physics uniquely enabled with polarized electron beams includes precision measurements of the tau EDM and tau Michel parameters. In addition, searches for lepton flavour violation in tau decays and dynamical mass generation hadronization studies will be enhanced with polarized beams~\cite{ChiralBelleWP}.

\setboolean{inbibliography}{true}
\addcontentsline{toc}{section}{References}
\bibliographystyle{B2Snowmass_bibtex}
\bibliography{B2Snowmass_references}
\setboolean{inbibliography}{false}

\end{document}